\documentclass[12pt]{article}

\usepackage{amssymb,amsmath,mathrsfs,epstopdf,slashed,color}
\usepackage{ifpdf}
\ifpdf
\usepackage{graphicx}
\usepackage{hyperref}
\usepackage{cite}
\else
\usepackage[dvipdfmx]{graphicx}
\usepackage[dvipdfmx]{hyperref}
\usepackage[numbers]{natbib}
\usepackage{hypernat}
\fi

\allowdisplaybreaks[1]

\makeatletter

\@addtoreset{equation}{section}
\makeatother

\newcommand{\Vol}{\mathcal{V}}
\newcommand{\cF}{\mathcal{F}}
\newcommand{\cE}{\mathcal{E}}
\newcommand{\cL}{\mathcal{L}}
\newcommand{\cC}{\mathcal{C}}

\DeclareMathOperator*{\re}{Re \,}
\DeclareMathOperator*{\im}{Im \,}

\newcommand{\PP}{\mathbb{P}}

\begin{document}

\begin{titlepage}

\setcounter{page}{0}
  
\begin{flushright}
 \small
  \normalsize
  KEK-TH-1864 \\
  OUTP-15-23P
\end{flushright}

\vskip -0.0cm
\begin{center}

{\Large \textbf{De Sitter vacua from a D-term generated racetrack  \\ \vspace*{5mm}  potential
in hypersurface Calabi-Yau compactifications}}

\vskip 0.5cm
  
{\large Andreas P. Braun${}^{1,2}$, Markus Rummel${}^{1}$, Yoske Sumitomo${}^{3}$ \\ and Roberto Valandro${}^{4,5,6}$}
 
 \vskip 0.5cm

{\small
${^1}$Rudolph Peierls Centre for Theoretical Physics, University of Oxford,\\ 1 Keble Road, Oxford, OX1 3NP, United Kingdom\\
${^2}$Mathematical Institute, University of Oxford, Woodstock Road,\\
Oxford, OX2 6GG, United Kingdom \\
${}^3$ High Energy Accelerator Research Organization, KEK,\\ 1-1 Oho, Tsukuba, Ibaraki 305-0801 Japan\\
${}^4$ Dipartimento di Fisica dell'Universit\`a di Trieste, Strada Costiera 11, 34151 Trieste, Italy\\
${}^5$ INFN, Sezione di Trieste, Italy\\
${}^6$ ICTP, Strada Costiera 11, 34151 Trieste, Italy\\
}

 \vskip 0.2cm

{\footnotesize Email: \href{mailto: andreas.braun@physics.ox.ac.uk, markus.rummel@physics.ox.ac.uk, sumitomo@post.kek.jp, roberto.valandro@ts.infn.it}{andreas.braun at physics.ox.ac.uk, markus.rummel at physics.ox.ac.uk, \\ sumitomo at post.kek.jp, roberto.valandro at ts.infn.it}}

\vskip 0.8cm
  
\abstract{\normalsize

In \cite{Rummel:2014raa} a mechanism to fix the closed string moduli in a de Sitter minimum was proposed: a D-term potential generates a linear relation between the volumes of two rigid divisors which in turn produces at lower energies a race-track potential with de Sitter minima at exponentially large volume. In this paper, we systematically search for implementations of this mechanism among all toric Calabi-Yau hypersurfaces with $h^{1,1}\leq 4$ from the Kreuzer-Skarke list. For these, topological data can be computed explicitly allowing us to find the subset of three-folds which have two rigid toric divisors that do not intersect each other and that are orthogonal to $h^{1,1}-2$ independent four-cycles. These manifolds allow to find D7-brane configurations compatible with the de Sitter uplift mechanism and we find an abundance of consistent choices of D7-brane fluxes inducing D-terms leading to a de Sitter minimum. Finally, we work out a couple of models in detail, checking the global consistency conditions and computing the value of the potential at the minimum.

}
  
\vspace{0.0cm}
\begin{flushleft}
 \today
\end{flushleft}
 
\end{center}
\end{titlepage}

\setcounter{page}{1}
\setcounter{footnote}{0}

\tableofcontents

\parskip=5pt

\section{Introduction}

Recent observations strongly confirmed the existence of Dark Energy, necessary for the present accelerated expansion of the universe \cite{Riess:1998cb,Schmidt:1998ys,Bennett:2012zja,Planck:2015xua}.
Among several possibilities, a tiny positive cosmological constant is the prime candidate explaining the observational data. As a candidate for unifying particle physics and quantum gravity, string theory should be able to accommodate such a possibility.

In string theory, flux compactifications generate a potential stabilizing moduli fields \cite{Giddings:2001yu,Dasgupta:1999ss} and
the cosmological constant is obtained as the minimal value of this potential.
As there are many consistent choices of quantized fluxes for a chosen compactification manifold,
we have a huge number of minima, resulting in the string theory landscape (for a review see \cite{Douglas:2006es,Grana:2005jc,Blumenhagen:2006ci,Silverstein:2013wua,Quevedo:2014xia,Baumann:2014nda}).
A positive cosmological constant is challenging to realize in this landscape while minima with negative cosmological constants seem ubiquitous. 
Given a vacuum with negative cosmological constant, we may ask if there is a mechanism to uplift the corresponding minimum of the potential to a de-Sitter (dS) solution, while keeping the stability of moduli fields. 

In the context of type IIB string theory, where moduli stabilisation has been extensively studied in the last fifteen years, 
several possibilities have been proposed to uplift the minima of the potential: anti-D3-brane uplift (KKLT) \cite{Kachru:2002gs,Kachru:2003aw,Kachru:2003sx}, non-SUSY stabilization of complex structure moduli \cite{Saltman:2004jh}, K\"ahler uplifting scenario \cite{Balasubramanian:2004uy,Westphal:2006tn,Rummel:2011cd,deAlwis:2011dp,Sumitomo:2013vla}, negative curvature \cite{Silverstein:2007ac}, D-term uplift \cite{Burgess:2003ic,Cremades:2007ig,Krippendorf:2009zza,Cicoli:2011yh}, and dilaton-dependent non-perturbative effects \cite{Cicoli:2012fh}.
These mechanisms consist in introducing new ingredients in the compactification, that modify the moduli potential by a positive definite contribution. 
In many of them the uplift potential needs to be tuned (by either warping, tuning of other fluxes, or coefficients accompanied by loop corrections) in order not to generate a runaway potential when added to the moduli stabilizing term.
Once the uplift term is added to the scalar potential, one minimizes the new potential and finds a dS minimum. Most of these mechanisms work at the level of the effective field theory (EFT) and so far have no intrinsic ten-dimensional (10D) description. Moreover they are based on quantum corrections that are not completely under control. This led many authors to consider also classical dS solutions realized at the level of the 10D theory \cite{Hertzberg:2007wc,Haque:2008jz,
Danielsson:2009ff,deCarlos:2009fq,
Wrase:2010ew,Danielsson:2010bc,Andriot:2010ju,Danielsson:2011au, Shiu:2011zt, VanRiet:2011yc, Danielsson:2012et,
Blaback:2013ht,
Damian:2013dwa}. The difficulty in finding these solutions, compared to the 4D ones, gave rise to some criticism on the 4D EFT approach and it opened the debate on the validity of KKLT anti-D3-brane uplift mechanism \cite{DeWolfe:2008zy,McGuirk:2009xx,Bena:2009xk}
(see \cite{Bena:2014jaa,Kallosh:2014wsa,Michel:2014lva,Hartnett:2015oda,Bergshoeff:2015jxa,Bena:2015kia,Cohen-Maldonado:2015ssa,Kallosh:2015nia,Polchinski:2015bea} for recent development). It is certainly an important point to understand the 4D EFT uplift mechanism from a 10D point of view, in order to be sure that they can be embedded into string theory.

Recently, another uplift mechanism has been proposed in the context of type IIB orientifold compactifications, studied by use of the 4D EFT language: the D-term generated racetrack uplift \cite{Rummel:2014raa}. A D-term generated by magnetized D7-branes forces a relation between the K\"ahler moduli and at lower energies induces a racetrack potential. The uplift term naturally chases after (or balances with) the stabilization potential through the dynamics without special suppressed coefficients.
In the Large Volume Scenario (LVS) \cite{Balasubramanian:2005zx}, the stabilization potential for the K\"ahler moduli includes a term proportional to $e^{-a_1 \tau_1}/\Vol^2$ where $\tau_1$ is the volume of a shrinkable four-cycle $D_1$ in the Calabi-Yau (CY) three-fold, generated by a non-perturbative contribution to the superpotential. $a_1$ is a coefficient which depends on how the non-perturbative effect is generated and $\Vol$ is the overall volume of the CY.
A second term like this is present in case another non-perturbative effect is generated on a second four-cycle $D_2$. Under the assumption that the contributions from the VEVs (vacuum expectation values) of matter fields to the D-term potential is fixed to zero by the open string moduli potential, the vanishing of the FI-term will force its volume $\tau_2$ to be proportional to $\tau_1$, such that $a_2 \tau_2 = \beta a_1 \tau_1$ with $\beta$ a real constant. In the effective potential we then have a second term that goes like $e^{-\beta a_1\tau_1}/\Vol^2$. This term contributes effectively to the uplift when $\beta<1$ in the D-term generated racetrack model. When the value of $\beta$ is closer to one, say $\beta \sim 0.9$, the racetrack potential has a better chasing, resulting in almost no tuning of the flux dependent parameters in the effective potential. 
The possible values of $\beta$ are model dependent: they depend on how the non-perturbative effects are generated, on the actual flux on the D7-brane and on the topological data of the Calabi-Yau three-fold
used for compactification. Note that the racetrack potential generated by the D-term constraint has also been applied to construct an inflationary model recently, alleviating the known concern of dangerous string-loop corrections \cite{Maharana:2015saa}.

In this paper, we will scan over a number of different models and explore which values for $\beta$ can be realized and how close we can get to $\beta \sim 1$. 
We work in a compact setup, where the global consistency conditions, such as tadpole cancellation, can be analysed in detail. For other works studying global dS vacua in type IIB compactifications while employing other mechanisms for the uplift see \cite{Cicoli:2012vw,Louis:2012nb,Cicoli:2013mpa,Cicoli:2013cha}.
We will first see what are the topological conditions a CY threefold must satisfy in order to host the necessary ingredients for the uplift mechanism to work: the main constraint is finding two rigid, shrinkable divisors which do not intersect each other and are furthermore orthogonal to $h^{1,1}-2$ linearly independent divisors.

A natural starting point to study when these constraints can be realized is given by Calabi-Yau threefolds which are hypersurfaces in toric varieties. Such manifolds can be constructed via combinatorial objects called reflexive polytopes \cite{Batyrev1993}, and the four-dimensional polytopes relevant to Calabi-Yau threefolds were 
famously classified in \cite{Kreuzer:2000xy}. In order to describe the topology of divisors and their intersections, we need slightly more refined data, which is given by appropriate triangulations of the polytopes found in \cite{Kreuzer:2000xy}. The task of computing all inequivalent triangulations was recently accomplished for small $h^{1,1}$ by \cite{Altman:2014bfa}, which serves as the input for our scan. Using their data, 
we compile a list of CYs with $h^{1,1} = 3,4$ which fulfil the necessary conditions. We find that roughly $10\%$ of polytopes have triangulations (often more than one) such that our conditions are met.
This list will be useful for other purposes and is a by-product of our work. 
In particular, these CYs can easily have two non-perturbative effects that contribute to the superpotential and on which one can play to stabilize more than two moduli in the Large Volume Scenario.
For each CY in the list, we consider the two rigid divisors, corresponding to K\"ahler moduli fields $T_1, T_2$; instantonic D3-branes wrapping these divisors will generate non-perturbative terms in the superpotential. 
For a specific class of orientifold involutions and D7-brane configurations, we scan over fluxes on the D7-branes generating a D-term potential. Correspondingly, we get a scan over the proportionality factor between $\tau_1$ and $\tau_2$ and in particular of possible values for $\beta$. For the classes of models we consider, we can quite easily obtain values of $\beta$ close to one.

We consider four classes of models: (a) rank one E3-instantons and four D7 wrapping the location of O7 plane $D_{\rm O7}$, (b) rank one E3-instantons and one D7 wrapping four-times of O7 divisor $4D_{\rm O7}$, (c,d) rank two instantons generating non-perturbative effects with both kinds of D7-brane configurations.
When we scan all possible models for $h^{1,1} = 3,4$, we find that class (a) has a maximal value of $\beta = 2/3$ existing only for $h^{1,1}=4$, and that the other three cases (b,c,d) can have $\beta = 8/9$ even at $h^{1,1} =3$, suggesting almost no special tuning of the coefficients towards realistic dS vacua. We hence conclude that the D-term generated racetrack uplift model is quite promising and compatible with a large number of CY compactifications.
We study three concrete examples and describe the data of the CY threefold, make a choice for the flux that maximizes the value of $\beta\lesssim 1$ and compute the actual values of the stabilized K\"ahler moduli, showing that the minimum of the scalar potential sits at a dS minimum.

In this paper we consider only setups with an orientifold involution where $h^{1,1}_-=0$. As explained in \cite{Blumenhagen:2007sm}, in these cases the flux on the D7-brane necessarily generates a charge for the instanton. Correspondingly, the prefactor of the non-perturbative superpotential is proportional to the VEVs of some D7-branes matter fields. Besides having an instanton zero mode structure that allows for non-zero Grassmanian integration in the instanton path-integral, one hence needs to have non-zero VEVs for the appearing matter fields in order to have non-zero contributions to the non-perturbative superpotential. For this reason, we computed the number of instanton zero modes and matter fields (for both chiralities) and check that there is no obstruction to a generation of the non-perturbative superpotential in the examples studied in the last part of the paper. This computation (which is reported in the appendix) can be avoided when $h^{1,1}_-\neq 0$ \cite{Grimm:2011dj}, as explained in Section \ref{Sec:discussion}. In this case, one can have a matter field independent prefactor of the non-perturbative superpotential.
Most of the analysis in the generic cases does not rely on the particular orientifold involution. Hence we believe that in constructions with $h^{1,1}_-\neq 0$, one will have the same abundance of order one values for $\beta$. We leave the analysis of these more generic cases, as well as the analysis of CYs with higher $h^{1,1}$, for future work.

This paper is organized as follows.
In Section \ref{Sec:ModStabIIB} we review type IIB orientifold compactifications with D7-branes and the stabilization of geometric moduli. In Section \ref{Sec:RevMechanism} we present the dS uplift mechanism, listing which conditions the global setup must satisfy to allow it. In Section \ref{Sec:SearchToric} we explain how to implement the search in the given a list of hypersurfaces in toric ambient spaces (i.e. triangulations of four-dimensional reflexive polytopes). In Section \ref{Sec:SearchFlux} we present the scan among CYs and D7-brane fluxes, yielding the possible values of $\beta$. In Section \ref{Sec:Examples} we present some concrete examples for specific choices of CYs and fluxes, and work out the details of the corresponding models. In Section \ref{Sec:discussion}, we summarize our results and discuss which features can be improved in the near future. Appendix \ref{sect:zeromodesdetails} contains the computation of the number of zero modes originating from D7-branes and their intersections (chiral as well as vector like pairs).

\section{Type IIB orientifold compactifications}\label{Sec:ModStabIIB}

We consider Type IIB string theory compactified on a Calabi-Yau three-fold $X_3$ with an orientifold involution. The effective four dimensional theory has $\mathcal{N}=1$ supersymmetry. The involution is chosen such that the fixed point set is made up of (complex) codimension 1 and codimension 3 objects, i.e. O7-planes and O3-planes. Moreover, it divides the homology classes of $X_3$ into an even and odd part: accordingly $h^{p,q}_{+(-)}$ denote the dimensions of the cohomology groups of even (odd) (p,q)-forms.

This compactification has the following set of massless fields coming from the closed string sector:
\begin{itemize}
\item the axio-dilaton $S=e^{-\phi}+i\,C_0$, where $\phi$ is the dilaton field and $C_0$ the RR scalar.
\item geometric moduli, which are divided into complex structure moduli $U_\alpha$ with $\alpha=1,...,h^{1,2}_-(X_3)$ 
and K\"ahler moduli $T_i=\tau_i+i \theta_i $ with $i=1,...,h^{1,1}_+$, where $\tau_i$ are the volumes $h^{1,1}_+$ independent even divisors $D^+_i$ and $\theta_i $ are KK zero modes scalars of the RR four-form potential $C_4$: $T_i=\int_{D^+_i} \tfrac12 J^2 + i C_4$.
\item G-moduli $G_k=C_k-iS\zeta_k$ with $k=1,...,h^{1,1}_-$ coming from KK zero mode scalars of the RR and NSNS two-forms $C_2$ and $B_2$.
\item KK zero mode vectors of the RR four-form $C_4$, $A_\mu^p$ with $p=1,...,h^{1,2}_+$.
\end{itemize}

There are also massless modes coming from the open string sector, i.e. from the D-branes. These are necessarily present in the given setup, 
as the RR charge of the O-planes must be cancelled to prevent a tadpole.

\subsection{D-brane configuration}\label{Sec:DbraneConfig}

In any consistent compactification of type IIB string theory, the total 7-brane charge must be zero on a compact manifold. This means that the 7-brane charge of the orientifold 7-planes
which are considering here must be cancelled by D7-branes, i.e.
\begin{equation}\label{TadpoleD7}
 \sum_A [D7_A] - 8 [O7] = 0 \:,
\end{equation}
where $A$ runs over all the D7-branes (and their orientifold images), $[D7_A]$ is the homology class of the divisor wrapped by the D7-brane $D7_A$ and $[O7]$ is the homology class of the fixed point locus. 
Furthermore, the D7-branes must be placed such that there is a double intersection between the orientifold plane and the D7-brane locus \cite{Collinucci:2008pf,Braun:2008ua}.
When the orientifold involution is realized by $\xi\mapsto -\xi$, these conditions are satisfied if the D7-brane locus is described by an equation of the form
\begin{equation}
  P_{D7} \equiv \eta^2 - \xi^2 \chi = 0  \:.
\end{equation}
The degrees of the polynomials (corresponding divisor classes) are dictated by the tadpole cancellation condition \eqref{TadpoleD7}. The orientifold locus $\xi=0$ is in the class $[O7]=[\xi]$, and hence the classes relative to $\eta$ and $\chi$ are $[\eta]=4[\xi]$ and $[\chi]=6[\xi]$.
If $\eta$ and $\chi$ are generic polynomials, the D7-brane locus is connected and hence we have only one orientifold invariant D7-brane. To have different branes, $\eta$ and $\chi$ must be restricted in a way that makes $P_{D7}$ factorize. The most trivial example is when $\eta\equiv \alpha \xi^4$ and $\chi\equiv \beta \xi^6$ with $\alpha$ and $\beta$ two different numbers. In this case
\begin{equation}
 P_{D7}\equiv (\alpha^2-\beta) \xi^8 =0  \qquad \qquad \mbox{with}\qquad \beta\neq \alpha^2\: ,
\end{equation}
and we have four D7-branes plus their four images all on top of the orientifold locus, realizing an $SO(8)$ gauge group. This will be the simple setup that we consider in the examples in Section \ref{Sec:Examples}.\footnote{If one of the non-perturbative effects needs to be generated via D7-brane gaugino condensation (that could be necessary to make the mechanism work) 
this additional D7 brane stack has to be consistent with tadpole cancellation as well.}
Another possible simple choice is to take $\eta$ generic and set $\chi\equiv \psi^2$, so that 
\begin{equation}\label{braneImagebrane}
P_{D7} \equiv \eta^2-\xi^2\psi^2 = (\eta - \xi\,\psi)(\eta + \xi\,\psi) =0 \,.
\end{equation}
This describes a D7-brane wrapping the locus $\eta - \xi\,\psi=0$ and its image at $\eta + \xi\,\psi$.
This is a second possibility that we will explore. 
In this case, the D7-brane and its image each wrap a divisor in the homology class $4D_\xi = 4[O7]$.

There is another important consistency conditions that the D7-brane configuration needs to satisfy. In order to prevent a Freed-Witten anomaly, the pull-back of the field strength of the B-field, $H_3=dB$, to the worldvolume of the D7-brane must be zero (if there are no branes ending on other branes), i.e.
\begin{equation}\label{H3FWcanc}
   i^\ast H_3 =0 \:,
\end{equation}
where $i^\ast$ is the pull-back map from the target space to the D7-brane worldvolume. This condition is obviously realized when the D7-brane worldvolume has no closed three-forms, or in other words when $h^{1,0}(D7)=0$. In the examples we provide in Section \ref{Sec:Examples} we will make sure to exclusively work with such branes.

A D7-brane may also support a non-trivial gauge bundle. The corresponding field strength $F$ (called also the gauge flux) 
must satisfy a proper quantization condition in order to cancel a second Freed-Witten anomaly:
\begin{equation} \label{FWanomalyCancel}
 F + \frac{c_1(D)}{2} = F - \frac{D}{2} \in H^2(D,\mathbb{Z}) \:,
\end{equation}
where we have used that on a Calabi-Yau manifold the first Chern class of any divisor is equal to minus the first Chern class of its normal bundle ($c_1(D)= - c_1(N_D) = -D$). When $D$ is not spin $\frac{c_1(D)}{2}$ is half-integral, so that $F$ must be half-integral as well (and in particular non-vanishing).
The gauge flux $F$ by itself is not invariant under the shift of the NS-NS two-form potential, the B-field. The gauge invariant combination is $\cF \equiv F - i^\ast B$, where $i^\ast B$ is the pull-back of $B$ on the brane worldvolume.

\subsubsection*{RR charges}

D7-branes wrapping a compact surface $D$ in $X_3$ supporting a holomorphic gauge bundle $ {\cal E} $ give rise to RR-charges of lower degrees, i.e. D5 and D3 charges, as well. These charges are conveniently presented as the Mukai vector, that is the polyform
\begin{align}\label{MukaiVect}
 \Gamma_{{\cal E}}  = e^{-B} D \wedge ch\left( {\cal E}\right) \wedge\sqrt{ \frac{Td\left( T D \right)}{Td \left( N  D\right)}}\,\, ,
\end{align}
where we are using the same symbol for the surface $D$ and its Poincar\'e dual two-form.

The polyform \eqref{MukaiVect} appears in the Chern-Simons action of the D7-brane, from which the charges with respect to the RR potentials can be simply
read off. Using the polyform $C=\sum_p C_p$ for all of the RR potentials, the Chern-Simons action is
\begin{equation}
\label{ChSimD7Action}
 S_{CS} = \int_{\mathbb{R}^{1,3}\times D} \iota^\ast C \wedge  \Gamma_{{\cal E}} \:.
\end{equation}
Keeping only the integral of the six-form, one can see that the 2-form in \eqref{MukaiVect} measures the D7-brane charge (it is Poincar\'e dual to the surface wrapped by the D7-brane), the 4-form counts the D5-brane charge, and the 6-form counts (minus) the D3-brane charge.
In particular, for an orientifold invariant D7-brane configuration in an orientifold compactification with $h^{1,1}_-(X_3)=0$, the total D5-charge will automatically cancel. 

For the D3-charge of a single D7-brane, one obtains from \eqref{MukaiVect}:
\begin{eqnarray}\label{QD3D7}
  Q_{D3}^{D7} &=& -  \int_{X_3} \Gamma_{{\cal E}} |_{\rm 6-form}  
  					  =  - \frac{\chi(D)}{24} - \frac12 \int_D \cF \wedge \cF \:,
\end{eqnarray}
where $\cF$ is the gauge invariant combination $F - i^\ast B$ and $F=c_1({\cal E}) + \frac{c_1(D)}{2}$. One can see that this is properly quantized when $\mathcal{E}$ is a well defined line bundle (we are considering a single D7-brane), i.e. $c_1({\cal E})$ is an integral two-form. Note that the Euler characteristic of a divisor $D$ of a Calabi-Yau threefold can be easily computed as $\chi(D)=\int_{X_3}(D^3+D\wedge c_2(X_3))$ by adjunction.

There is an analogous expression for the charges induced by the O7-plane wrapping $D_\xi$. Its Mukai vector is simply given by
\begin{equation}
 \Gamma_{O7} = - 8 D_{\xi} + \frac{\chi(D_{\xi})}{6} \:,
\end{equation}
from which the D7-brane charge and the D3-brane charge of an O7-plane can be read off analogously. Cancellation of the D7-brane tadpole
gives rise to the relation \eqref{TadpoleD7}.

Let us now consider a configuration with one O7-plane and four D7-branes plus their images wrapping the same divisor $D_\xi$. We choose a diagonal gauge flux along the four D7-branes (i.e. each D7-brane support the same line bundle) and the proper flux on their images to make the configuration invariant (i.e. $\cF'=-\sigma^\ast \cF$). Then the total D3-charge is given by 
\begin{equation}\label{O74D7QD3}
 Q_{D3}^{O7+4D7} =  - \frac{\chi(D_\xi)}{2} - 4 \int_{D_\xi} \cF \wedge \cF \:.
\end{equation}

For the configuration \eqref{braneImagebrane}, where we have one D7-brane wrapping a divisor in the class $4D_\xi$ plus its image, 
the D3-charge contribution is
\begin{equation}\label{O71D7QD3}
 Q_{D3}^{O7+1D7} =  - \frac{\chi(D_\xi)}{6} - \frac{\chi(4D_\xi)}{12} -  \int_{4D_\xi} \cF \wedge \cF \: ,
\end{equation}
where $\cF$ denotes the flux on the D7-brane and $-\cF$ the flux on its orientifold image (we are assuming $h^{1,1}_-(X_3)=0$). 
Notice that the geometric contribution is now larger than the case with four branes wrapping $D_\xi$, as $\chi(nD)$ goes roughly like $n^3\chi(D)$. 

Finally, there is also a contribution to the D3-brane charge coming from O3-planes, i.e. fixed points of the orientifold involution.
Every such fixed point contributes $-1/2$ to the total D3-charge.

A consistent D-brane background also needs to have total zero K-theoretic torsion charges (that is not captured by \eqref{MukaiVect}). In the setup that we will consider, we check that global $SU(2)$ Witten anomaly are cancelled in the worldvolume of every invariant $Sp(1)$ probe brane transverse to the O-plane, which is an equivalent condition \cite{Uranga:2000xp}. This anomaly is cancelled if each intersection of the probe branes with the D7-branes in the setup support an even number of fields in the fundamental representation of the probe group. For the simple configurations that we will consider in this paper, this always happens.\footnote{Either we consider an even number of D7-branes (that realize a even rank flavour group) or they will wrap an even cycle, making the number of chiral fields even.}

\subsubsection*{Zero modes}

The constructions we are interested in contain brane configurations made up of D7-branes and D3-branes wrapping four-cycles of the compact Calabi-Yau three-fold $X_3$. The D3-branes are point-like in the non-compact space-time directions and are called E3-instantons (`E' stays for Euclidean). 

There are massless open strings living at each intersection of D7-branes with E3-branes. These are the zero modes of the stringy instanton background (the D7-branes are seen as the flavour group of the zero modes). Having control over the instanton fermionic zero modes is important to understand whether the E3-brane will contribute to the non-perturbative superpotential. The relevant E3-branes will wrap rigid divisors with $h^{1,0}=0$, so that they have the minimal number of neutral zero modes, i.e. two states. Furthermore, E3-branes will generally intersect some D7-branes (to allow the uplift mechanism) on a curve $\mathcal{C}$. At such intersections, there are fermionic zero modes in the bifundamental representation (w.r.t. the gauge groups on the E3- and on the D7-brane stacks). If we have a single E3-brane with flux $F_{E3}=c_1(E_{E3})$, $E_{E3}=\cE_{E3}\otimes K_{E3}^{-1/2}$ as well as $n_D$ D7-branes with a diagonal flux $F_D=c_1(E_{D})$, $E_{D}=\cE_{D}\otimes K_{D}^{-1/2}$, the zero modes are counted 
by the following extension groups
\begin{eqnarray}\label{ExtE3D7}
 \mbox{Ext}^1\left(i_\ast E_{E3}, i_\ast E_D\right) &=& H^1\left(\mathcal{C},E_{E3}^\vee\otimes E_{D} \otimes K_{\mathcal{C}}^{1/2} \right)  \:,\\
  \mbox{Ext}^2\left(i_\ast E_{E3}, i_\ast E_D\right) &=& H^0\left(\mathcal{C},E_{E3}^\vee\otimes E_{D} \otimes K_{\mathcal{C}}^{1/2} \right) \:,
\end{eqnarray}
where $K_{\mathcal{C}}=D_{E3}+D_{D}$ is the canonical bundle of the intersection curve. The states counted by $\mbox{Ext}^1$ are in the fundamental representation ${\bf n_D}$ of the D7-gauge group, while the ones counted by $\mbox{Ext}^2$ are in anti-fundamental representation ${\bf \bar{n}_D}$.
The difference between their numbers is given by the chiral index
\begin{eqnarray}\label{IndexD7E3}
 I_{D7-E3} &=& \dim \mbox{Ext}^2\left(i_\ast E_{E3}, i_\ast E_D\right) - \dim \mbox{Ext}^1\left(i_\ast E_{E3}, i_\ast E_D\right)  = \chi\left(\mathcal{C},E_{E3}^\vee\otimes E_{D} \otimes K_{\mathcal{C}}^{1/2} \right) \nonumber \\
     &=& \int_{\mathcal{C}} F_{D} - F_{E3} = \int_{\mathcal{C}} \cF_{D} - \cF_{E3} \:,
\end{eqnarray}
where $ \chi\left(\mathcal{C},E_{E3}^\vee\otimes E_{D} \otimes K_{\mathcal{C}}^{1/2} \right)$ is the holomorphic twisted Euler characteristic. In the second line we have used the Hirzebruch-Riemann-Roch theorem.\footnote{The theorem says that $\chi(M,E)=\int_M \mbox{ch}(E)\,\,\mbox{Td}(M)$, where $\chi(M,E)\equiv \sum_n (-1)^n h^n(M,E)$, $M$ is a manifold of any dimension and $E$ is a holomorphic bundle defined over the manifold.}. In the models in Section \ref{Sec:Examples}, we will compute the dimension of both groups in \eqref{ExtE3D7}, because we need to know the number of all the zero modes.

Massless open string states living on the worldvolume of a D7-brane or intersections of D7-branes are seen at low energies as massless four-dimensional fields.
Let us first discuss the case of a D7-brane and its image, described by \eqref{braneImagebrane}. In this case the computation of the zero modes is complicated by the presence of the O7-plane. Luckily, this situation has been studied in many cases in the literature \cite{Collinucci:2008pf,Collinucci:2014taa,Braun:2011zm,Collinucci:2008sq}. One can describe this system by a tachyon matrix\footnote{This tachyon is the field which makes a $D9-\bar{D9}$ system condense to the $7$-brane configuration we are interested in, see \cite{Collinucci:2008pf} for the details.}
\begin{equation}
 T\,:\qquad  [\cF_D]^{-1} [D_\xi]^{-2} \oplus    [\cF_D]^{-1} [D_\xi]^{2}  \qquad \rightarrow \qquad  [\cF_D]\, [D_\xi]^{2} \oplus    [\cF_D]\, [D_\xi]^{-2}  \:,
\end{equation}
with 
\begin{equation}
 T \, = \, \left(\begin{array}{cc}
0 & -\eta-\xi\psi \\  \eta - \xi\psi & 0 \\ 
\end{array}   \right)\,.
\end{equation}
The two branes intersect in two loci: $\mathcal{C}=\{\eta=\psi=0\}$ and $\mathcal{C}_0=\{\eta=\xi=0\}$. Only the first curve produces charged fields (in the symmetric representation of $U(1)$), while the second curve is empty (the antisymmetric representation of $U(1)$ is trivial).
The zero modes are the deformations of this matrix, up to linearised gauge transformations \cite{Collinucci:2008pf}. The diagonal terms (which must be proportional to $\xi$ to make the Tachyon matrix orientifold invariant) give the states localized on the curve $\cC=\{\eta=\psi=0\}$: $\delta T_{11}=\xi\delta\rho$ are states of charge $+2$, while $\delta T_{22}=\xi\delta\tau$ have charge $-2$. The matrix $\delta T$ is well defined if $[\delta\rho]=[\cF]^2[D_\xi]^3$ and $[\delta\tau]=[\cF]^{-2}[D_\xi]^3$ (where the subtraction of $[D_\xi]$ has been taken into account). Hence the numbers of these states are counted respectively by
\begin{eqnarray}\label{ExtD7D7image}
H^0\left(\mathcal{C},[\cF_D]^2 \otimes [D_\xi]^3 \right)  & \mbox{ and } & 
H^0\left(\mathcal{C},[\cF_D]^{-2} \otimes [D_\xi]^3 \right) \: ,
\end{eqnarray}
and their difference is given by the index
\begin{eqnarray}
 I_{D7D7'} &=& \frac12\int_{X_3} D_D^2(\cF_D-\cF'_D)+ D\,D_\xi\cF_D
 = 12\int_{X_3} D_\xi \wedge D_\xi \wedge \cF_D   = \int_{\mathcal{C}} 2 \cF_{D}\:,
\end{eqnarray}
where we have used $D=4D_\xi$.

Let us now consider a case where we have two stacks of D7-branes wrapping the same divisor $D$, one supporting an overall line bundle $E_a=\cE_a\otimes K_{D}^{-1/2}$ and the other the line bundle $E_b=\cE_b\otimes K_{D}^{-1/2}$. Then the massless spectrum is counted by Ext$^n(i_\ast E_a, i_\ast E_b)$ with $n=0,...,3$, where 
\begin{eqnarray}\label{StatesMagnBr}
 \mbox{Ext}^0\left(i_\ast E_{a}, i_\ast E_b\right) &=& H^0\left(D,E_{a}\otimes E_{b}^\vee \right)  \:,\\
 \mbox{Ext}^1\left(i_\ast E_{a}, i_\ast E_b\right) &=& H^1\left(D,E_{a}\otimes E_{b}^\vee\right)+H^0\left(D,E_{a}\otimes E_{b}^\vee \otimes N_{D} \right)   \:, \\
 \mbox{Ext}^2\left(i_\ast E_{a}, i_\ast E_b\right) &=& H^2\left(D,E_{a}\otimes E_{b}^\vee\right)+H^1\left(D,E_{a}\otimes E_{b}^\vee \otimes N_{D} \right)   \:, \\
 \mbox{Ext}^3\left(i_\ast E_{a}, i_\ast E_b\right) &=&H^2\left(D,E_{a}\otimes E_{b}^\vee \otimes N_{D} \right)   \:.
\end{eqnarray}
One can relate some of the cohomology groups by using Serre duality, $H^i\left(D,E_{a}\otimes E_{b}^\vee\otimes N_D\right)=H^{2-i}\left(D,E_{a}^\vee\otimes E_{b} \right)$. The states counted by the extension groups with 
$n=0$ and $n=3$ are ghosts that need to be absent for the consistency of the D-brane configuration \cite{Donagi:2008ca,Blumenhagen:2008zz}. These groups are empty when $h^0\left(D,E_{a}\otimes E_{b}^\vee\right) =h^0\left(D,E_{a}^\vee\otimes E_{b} \right) =0$. As shown in \cite{Blumenhagen:2008zz}, this is always true if there exists a value of the K\"ahler form $J$ inside the K\"ahler cone, for which the bundle choice is supersymmetric. The states corresponding to $n=2$ are bifundamentals in the $({\bf \bar{n}_a},{\bf n_b})$ representation, while those with $n=1$ are in the conjugate representation.
The difference between their numbers is given by the chiral index
\begin{eqnarray}
 I_{ab} &=& \dim \mbox{Ext}^2\left(i_\ast E_{a}, i_\ast E_b\right) - \dim \mbox{Ext}^1\left(i_\ast E_{a}, i_\ast E_b\right) \\  
     &=& \chi\left(D,E_{a}\otimes E_{b}^\vee \right) - \chi\left(D,E_{a}\otimes E_{b}^\vee \otimes N_{D} \right) \\
     &=& \int_{X_3} D\wedge D\wedge (F_{b} - F_{a}) = \int_{X_3} D\wedge D\wedge (\cF_{b} - \cF_{a}) \:.
\end{eqnarray}
The case we are interested in is where all D7-branes are on top of the orientifold brane, so that the two stacks are the orientifold
images of one another. This gives rise to the relation $E_b=\sigma^\ast E_a^\vee $ between the bundles on their worldvolumes.

\subsection{Moduli stabilization}

Type IIB string theory compactified on a CY orientifold with a given configuration of D-branes has a plethora of massless scalars in the low energy effective theory, the {\it moduli} of the compactification. These scalars need to be stabilized. We hence must introduce further ingredients to make this happen, i.e. study a slightly more complicated background. This problem has been addressed in the last fifteen years: the complex structure moduli and the axio-dilaton are stabilized at tree level by non-zero expectation values for the NSNS and RR three-form field strengths $H_3=dB_2$ and $F_3=dC_2$.  At subleading order, the K\"ahler moduli are fixed by non-perturbative corrections to the superpotential and perturbative corrections to the K\"ahler potential. There are two main scenarios in which this can happen. In the case of KKLT vacua \cite{Kachru:2003aw} the non-perturbative effects are enough to stabilise the K\"ahler moduli. On the other hand, if one uses a combination of both corrections to the tree level potential, that leads to the Large Volume Scenario (LVS) \cite{Balasubramanian:2004uy,Balasubramanian:2005zx} where the volume modulus is stabilized at exponentially large values.  In presence of magnetized D7-branes, some combination of the K\"ahler moduli can be stabilised already at tree-level by a D-term potential.

\subsubsection*{Effective potential}

In $\mathcal{N}=1$ supergravity, the scalar potential is a sum of the F-term and the D-term potentials. 

The F-term potential is completely determined by the K\"ahler potential $K$ and the superpotential $W$:
\begin{equation}\label{N1ScalarPot}
 V_F = e^K\left(  K^{A\bar{B}} D_AW \overline{D_BW} -3 |W|^2\right) \:,
\end{equation}
where $D_AW=\partial_A W+K_A W$ and $K_{A,...}$ are the derivatives of $K$ with respect to the scalars $\phi_A,...$. In our case, $A$ runs over the complex structure moduli $U_\alpha$ $\alpha=1,...,h^{1,2}_-$, the axio-dilaton $S$ and the complexified K\"ahler moduli $T_i$ $i=1,...,h^{1,1}_+$ (we neglect the moduli $G_k$ as we only consider models where $h^{1,1}_-=0$).

In Type IIB orientifold compactifications, the K\"ahler potential $K$ is
\begin{equation}\label{KpotIIB}
K = -2 \ln \left( \Vol + \frac{\hat{\xi}}{2} \right) - \ln \left( S+\bar{S}\right)  - \ln \left(i\int_{X_3} \Omega \wedge \bar{\Omega} \right) \:.
\end{equation}
Here, $S$ is the axio-dilaton, $\Omega$ is the holomorphic (3,0)-form on the CY three-fold $X_3$, which depends on the complex structure moduli $U_\alpha$, and $\Vol=\frac16\int_{X_3} J^3$ is the volume of $X_3$, which encodes the K\"ahler moduli $T_i$. In the K\"ahler potential of the K\"ahler moduli, we 
have also included the leading order $\alpha'$ corrections to the K\"ahler moduli, coming from compactifying the $\alpha'^3R^4$ term in the ten-dimensional effective theory. This produces a constant shift inside the log, where
\begin{equation}\label{AlphaPrCorrK}
  \hat{\xi} = - \frac{\zeta(3) \chi(X_3)}{4\sqrt{2}(2\pi)^3}(S+\bar{S})^{3/2} \:, 
\end{equation}
with $\chi(X_3)$ the Euler characteristic of the CY three-fold and $\zeta(3)\simeq 1.202$. Recently it has been claimed that the constant $\hat{\xi}$ is modified if one includes also the proper $\mathcal{N}=1$ contribution coming from the O7/D7 sector \cite{Minasian:2015bxa}.
This modification boils down to replacing $\chi (X_3)\mapsto \chi(X_3) + 2\int_{X_3}D_\xi^3$ in \eqref{AlphaPrCorrK}.\footnote{For the Large Volume Scenario to take place, one needs $\hat{\xi}>0$ that without the new term means $\chi(X_3)<0$. The second term could spoil this condition on $\chi(X_3)$ if $\int D_\xi^3>0$. This is actually positive in phenomenologically interesting models, as one usually chooses $D_\xi$ to be a large degree divisor (to produce a large negative D3-charge that would give large tunability on the flux choice) which hence has many effective representatives: this means that $D_\xi^3\geq0$. Luckily in all the explicit LVS-type models studied in the literature \cite{Cicoli:2011qg,Cicoli:2012vw,Cicoli:2013mpa,Cicoli:2013zha,Louis:2012nb,Cicoli:2013cha} and in the examples presented here, this term is not large enough to turn $\hat{\xi}<0$.} 

In Type IIB flux compactifications, there is a tree level superpotential $W^{\rm tree}$ generated by the three-form fluxes $H_3$ and $F_3$; this is the famous Gukov-Vafa-Witten superpotential \cite{Gukov:1999ya}
\begin{equation}\label{GVWsuperpot}
W^{\rm tree} = \int_{X_3} \left( F_3 -i\, S \, H_3 \right) \wedge \Omega \:.
\end{equation}
Furthermore, the superpotential can get subleading contributions at the non-perturbative level. These terms can be generated by E3-branes wrapping a divisor in $X_3$, or by gaugino condensation on some non-Abelian stacks of D7-branes \cite{Witten:1996bn,Kachru:2003aw}:
\begin{equation}
W^{\rm np} = \sum_j A_j e^{-a_j T_j} \:. \label{sumnonper}
\end{equation}
The prefactor $A_j$ depends generically on the complex structure moduli and the matter fields present in the model; the coefficient $a_j$ depends on the nature of the non-perturbative effect (for a rank-one E3-brane instanton, $a_j=2\pi$). 
The total superpotential in \eqref{N1ScalarPot} is the sum of the two terms: $ W= W^{\rm tree} + W^{\rm np}$.

When some D7-branes are present in the background, a D-term potential is generated. In the case of a $U(1)$ gauge field coming from D7-branes wrapping the divisor $D$ in $X_3$ (e.g. a single D7-brane of the diagonal $U(1)$ of a $U(N)$ stack) it is given by \cite{Haack:2006cy} :
\begin{equation}
 V_D = {1 \over \re (f_D)} \left(\sum_j c_{Dj} \hat{K}_j \phi_j - \zeta_D\right)^2, 
  \label{D-term potential}
\end{equation}
where $\phi_j$ are matter fields with charges $c_{Dj}$ under the $U(1)$ gauge symmetry, 
\begin{equation}\label{FItermD7}
 \zeta_D = {1\over {\cal V}} \int_{D_D} {\cal F}_D \wedge J \,,
\end{equation}
is the FI-term, which can be generated by the gauge flux $\cF_D$, and
\begin{equation}
 \re (f_D) = \frac12 \int_{D_D} J \wedge J - \frac{1}{2 g_s} \int_{D_D} \mathcal{F}_D \wedge \mathcal{F}_D   \,,
\end{equation}
is the gauge kinetic function.

We consider only vacua where the volume is stabilized at extremely large values ($\gtrsim 10^6 \ell_s^6$). We can hence expand the scalar potential in powers of $\frac{1}{\Vol}$ and write it as 
\begin{equation}
 V = V_F^{\rm tree} + V_D + V_F^{\rm sub-lead}\:.
\end{equation}
In a large volume expansion, one can fix the moduli order by order.

\subsubsection*{Complex structure moduli and axio-dilaton}

At leading order in the $1/\Vol$ expansion, the potential is given by \eqref{N1ScalarPot}, where $K$ is \eqref{KpotIIB} with $\hat{\xi}\rightarrow 0$ and $W=W^{\rm tree}$ \eqref{GVWsuperpot}. The tree-level superpotential depends on the integral of the three-forms $H_3$ and $F_3$ over a basis of three-cycles of $X_3$. Because of flux quantization, these are discretely tunable parameters in the effective four-dimensional theory.  They must satisfy the constraint coming from the tadpole cancellation condition: in fact, they induce a (typically positive) D3-charge that needs to be cancelled by (negative) contributions coming from D7-branes and O-planes:
\begin{equation}
\frac{1}{(2\pi)^4{\alpha'}^2} \int F_3\wedge H_3 \, + \,  Q_{D3}^{\rm loc} \,=\, 0 \:.
\end{equation}

Because of the no-scale property of the tree-level K\"ahler potential, the resulting potential is positive definite. Its only dependence on the K\"ahler moduli is a prefactor that goes like $1/\Vol^2$.
It is stabilized to zero value when $D_{U_\alpha}W=D_SW=0$ \cite{Giddings:2001yu}. The complex structure moduli and the axio-dilaton (in particular $g_s$) are fixed to three-form flux valued functions.
At this level, the K\"ahler moduli are flat directions. In the minimum, the tree-level superpotential is $\langle W^{\rm tree}\rangle = W_0$, where $W_0$ is a function depending on the flux quanta.\footnote{When D7-branes are present, there are open string moduli that describe the deformation of the D7-brane loci. Furthermore, there are gauge fluxes, which, if not of pull-back type, can stabilize some of the open string moduli \cite{Bianchi:2011qh,Bianchi:2012kt}. The proper language to describe this situation is F-theory: here, the complex structure moduli of $X_3$ and the open string moduli are both encoded in the holomorphic (4,0)-form of the Calabi-Yau four-fold, while the three form fluxes and the gauge fluxes are described by the periods of a four-form flux $G_4$. The GVW superpotential in this case is $W^{\rm tree}=\int_{X_4} G_4 \wedge \Omega_4$. One can make analogous consideration as above and see that the tree-level potential is again stabilised at zero value, keeping the K\"ahler moduli 
as flat directions. $W_0$ will now depend on the gauge fluxes as well. }

\subsubsection*{K\"ahler moduli: D-term}

The D-term potential \eqref{D-term potential} usually appears at $1/\Vol^2$ order as well. This is due to the fact that typically $\zeta_D\sim \Vol^{-2/3}$ and $\re(f_D)\sim \Vol^{2/3}$. It is also positive definite and is stabilized at zero value. This forces a relation between the VEV of the matter fields $\phi_j$ and the FI-term $\zeta_D$. If the matter field contribution is stabilized at zero, the D-term condition becomes 
\begin{equation}
 \zeta_D = 0  \qquad \Rightarrow \qquad \int_{D} J \wedge \cF_D =0\:.
\end{equation} 
If we expand the K\"ahler form in a basis of divisors $\{D_i\}$, $J=\sum_i t_i D_i$,  vanishing of the D-term is a linear equation on the coefficients of the K\"ahler form $t_i$. By inverting  
\begin{equation}
 \re T_i = \tau_i = \tfrac12 \int_{D_i} J^2 = \tfrac12 \kappa_{ijk}t^jt^k \:,
\end{equation}
one obtains relations among the K\"ahler moduli.

In the models we will consider in this paper, the FI-term will go like $\Vol^{-1}$ instead of $\Vol^{-2/3}$. Hence $V_D$ appears at $1/\Vol^{8/9}$, which is subleading with respect to $V_F^{\rm tree}$ but still leading with respect to $V_F^{\rm sub-lead}$ if the volume is very large.

\subsubsection*{K\"ahler moduli: F-term}

In the Large Volume Scenario, perturbative corrections to the K\"ahler potential and non-perturbative corrections to the superpotential, generate a scalar potential for the K\"ahler moduli (that are left unfixed by the leading terms) at order $1/\Vol^3$. Considering one non-perturbative object wrapping the divisor $D_s$, the scalar potential looks like
 \begin{equation}\label{SubLeadFtermPot}
V_F^{\rm sub-lead} \sim \left(\frac{A e^{-2a \tau}}{\mathcal{V}}-\frac{Be^{-a\tau}W_0}{\mathcal{V}^2}+\frac{C |W_0|^2}{\mathcal{V}^3}\right)\:.
\end{equation}
This potential is minimized when
\begin{equation}
\tau_s\sim 1/g_s \qquad \mbox{and} \qquad\mathcal{V}\sim W_0 \,e^{a_s\tau_s} \:.
\end{equation}
For small string coupling and $W_0$ of order one, the volume $\mathcal{V}$ is stabilized at an exponentially large value. If there are K\"ahler moduli that are left unfixed by this potential, one needs to consider sub-leading terms (like $g_s$-corrections to the K\"ahler potential).
The minimum of the potential is anti-de Sitter with broken supersymmetry.\footnote{In LVS,  $M_s\sim M_p/\mathcal{V}^{1/2}\gg M_{KK}\sim M_p/\mathcal{V}^{2/3}$ and both are much larger than the gravitino mass $m_{3/2}\sim M_p/\mathcal{V}$. Most moduli masses scale like the gravitino mass, except for the overall volume modulus itself which has a mass of order $m_{\mathcal{V}}\sim  M_p/\mathcal{V}^{3/2}\ll m_{3/2}$ and its axion partner which is essentially massless.} In this case, one needs to introduce new ingredients or go to less generic situations to end up with a de Sitter minimum.

\section{De Sitter vacua via a D-term generated racetrack}\label{Sec:RevMechanism}

In this section, we review the setup studied in \cite{Rummel:2014raa}, where a new mechanism for de Sitter uplift in Type IIB string vacua was introduced.

The basic ingredients are a couple of non-perturbative contributions to the superpotential (like the ones in \eqref{sumnonper}) and a D-term potential that depends on the volumes of both four cycles wrapped by the non-perturbative objects. At leading order in $1/\mathcal{V}$ the D-term vanishes, forcing the two volumes to be proportional to each other. At the next order in $1/\mathcal{V}$, the potential is given by the F-term of the K\"ahler moduli. The two non-perturbative contributions now generate a racetrack model that allows to have a de Sitter minima.

Let us see how this mechanism works in more detail.

\subsection{E3-instantons and non-perturbative superpotential}\label{E3instantons_sec}
The starting point is the presence in the superpotential of \textit{two non-perturbative contributions of the same order of magnitude}:
 \begin{equation}
  W = W_0 + A_1 e^{-a_1 T_1} + A_2 e^{-a_2 T_2}\,.\label{Wthreemod}
 \end{equation}
According to what we said in Section \ref{Sec:ModStabIIB}, these two terms will arise from two divisors $D_1$ and $D_2$ wrapped by E3 instantons, or a stack of D7-branes supporting gaugino condensation. In this paper we will restrict ourselves to study the contributions of the first type of non-perturbative contributions, i.e. E3 instantons wrapping orientifold invariant divisors. We furthermore require that $D_1$ and $D_2$ do not intersect each other.
These instantons contribute to the superpotential \eqref{Wthreemod} if the zero mode structure is constrained as follow: there are only two neutral zero modes -- this happens when the wrapped divisor is rigid and has $h^{1,0}=0$ -- and the path integral on the charged ones is non zero. This last condition can be satisfied if the VEV of some fields is non-zero, as we will explain later.

Another consistency condition that the E3 instantons have to satisfy is Freed-Witten anomaly cancellation. First they must satisfy \eqref{H3FWcanc}, which is realized when $h^{1,0}(D_i)=0$. Moreover, the gauge flux $F_i$ on the E3 brane wrapping the divisor $D_i$ is quantized according to \eqref{FWanomalyCancel}. Hence, when the divisor $D_i$ is non-spin, the gauge flux will always be different from zero.

Let us consider a rank one $O(1)$ instanton, i.e. one single D3-brane wrapping the divisor $D_i$.
Under the orientifold involution $\sigma$, the gauge invariant field strength $\cF_i=F_i-i^\ast B$ transforms as $\cF_i \mapsto - \sigma \cF_i$. If the involution is such that $h^{1,1}_-=0$, we have that $\sigma \cF_i=\cF_i$ and the E3-brane is orientifold invariant only if $\cF_i=0$. Hence, the non-perturbative effect survives the orientifold projection if $\cF_i=0$. When $i^\ast B=0$ and $D_i$ is non-spin, $\cF_i$ is always non-zero and then the E3 instanton is projected out. The E3 instanton is invariant when the background B-field is such that $\cF_i$ can be zero, i.e. $i^\ast B=F_i$. 
The choice $B = - D_1/2 - D_2/2$ will always allow to put $\cF_i=0$ when $D_1\cap D_2=0$.

One can have invariant E3 instantons wrapping non-spin divisor even when $B=0$. This is possible if we give up having rank one instantons, as introduced in \cite{Berglund:2012gr}. One E3 instanton is a stack of D3-branes wrapping the divisor $D_i$ in the CY $X_3$. As for the D7-branes, a vector bundle can live on such a stack. We take $D_i$ to be an invariant divisor transverse to the O7-plane. Hence, an E3 instanton with vector bundle $E$ is orientifold invariant if
\begin{equation}\label{InvE3cond}
  \sigma^\ast E^\vee \,=\, E \:.
\end{equation}
When $E$ is a line bundle (rank one instanton), this condition implies $c_1(E)=F=0$, which violates the Freed-Witten flux quantization for non-spin divisors. On the other hand, if $E={\cal L}_1\oplus {\cal L}_2$, i.e. it is a rank two vector bundle that is a sum of two line bundles, the condition \eqref{InvE3cond} boils down to requiring ${\cal L}_2={\cal L}_1^{-1}$. Hence this stack is made up of two branes, one is the image of the other, and $c_1({\cal L}_i)$ are the gauge fluxes on the branes, that can be non-zero without violating Freed-Witten anomaly cancellation.

The classical action for an E3 instanton wrapping the divisor $D$ is $S_{E3}= 2\pi \cdot $rk$(E)\cdot$vol$(D)$. Hence, while in \eqref{Wthreemod} $a_i=2\pi$ for a rank one E3-brane, for rank two instantons we have $a_i=4\pi$. If one divisor allows a rank one instanton, its contribution to the non-perturbative superpotential will always be dominant with respect to higher rank instantons; if it is not allowed, the leading contribution will be given by the rank two E3-brane (if non-zero).

\subsubsection*{Charged zero modes}

An important constraint one has to take into account is the general conflict between non-vanishing non-perturbative effects and fermionic zero modes at brane intersections~\cite{Blumenhagen:2007sm} generically generated by gauge fluxes. In our situation, we have some D7-branes that, as we discuss soon, will necessary intersect the E3 instantons and support a non-trivial flux (for the uplift mechanism to work). 
The effective superpotential generated by a E3 instanton is given by \cite{Witten:1996bn} (see also \cite{Martucci:2014ema})
\begin{equation}
 A \sim \int d\eta_1 \dots d\eta_n e^{-S_{\text{inst}}}\,,\label{Apathint}
\end{equation}
where $S_{\text{inst}}$ is the instanton effective action for $n$ fermions $\eta_i$. Their multiplicity is given by 
$n = n_D  \times n_f$, 
where $n_D$ is dimension of the representation of the D7-brane gauge group where the fermion lives (the number of D7-branes wrapping the divisor $D_D$)  and $n_f$ the number of solutions of the Dirac equation in the given representation. 
$S_{\text{inst}}$ includes all possible {\it gauge invariant} interaction terms between the $\eta_i$ and the scalar fields $\phi$ living on the D7-branes. 
Let us discuss with the case of one scalar field $\phi$ and an even or odd number $n$ of fermions $\eta_i$. The instanton action contains gauge invariant terms
$S_{\text{inst}} \supset \sum_{i,j=1}^n g_{ij} \eta_i \phi \eta_j$, 
with coupling constants $g_{ij}$. The integral in \eqref{Apathint} can be evaluated for an even $n$: one expands the exponential and keeps only terms that saturate the Grassmanian integration. 
On the other hand, for odd $n$ the contribution to the superpotential is zero \cite{Cvetic:2007ku,Blumenhagen:2007zk,Blumenhagen:2007bn,Cvetic:2007sj}. 

In the simple cases that we will consider, $n$ will always be even, because either $n_D$ or $n_f$ will be even. When we saturate the O7-plane charge by a stack of four D7-branes (plus their images), $n_D=4$. When the tadpole is cancelled by one brane (and its image), this brane wraps an even divisor ($[D7]=4[O7]$) and $n_f$ turns out to be even.
In these cases, the non-perturbative effects on $D_1$ and $D_2$ are never trivially zero but are proportional to a positive power of the scalar fields. Of course we need to check that the matter fields on the D7-branes have the right charges to produce the needed interaction terms. Finally, it is necessary for the non-perturbative effects not to vanish that the stabilization of the matter fields leads to non-zero VEVs $\langle \phi \rangle \neq 0$.\footnote{These are hidden sector fields so there are no constraints for standard model fields to obtain a VEV.} This is a strong assumption on the constructions presented in this paper. 
It could be made mild by considering a more involved setup, as we will discuss in Section \ref{Sec:discussion}.
This modification would not change the salient features regarding the dS uplift mechanism studied in this paper, while it would complicate the details. We hence leave these setups for a future work.

\subsection{D-term potential from D7-branes}

The other necessary ingredient is a stack of magnetized D7-branes generating a  D-term potential. 
We will call the divisor wrapped by such D7-branes $D_D$ and the gauge invariant worldvolume flux $\cF_D\equiv F_D - i^\ast B$. 

The FI-term \eqref{FItermD7} is proportional to a linear combination of the K\"ahler moduli $t_i$, where the $t_i$'s are the coefficients of the K\"ahler form expansion in terms of a basis of two forms. We will take the independent rigid divisors $D_1$ and $D_2$ among the basis elements:
\begin{equation}\label{Kahlerform}
 J = t_1 D_1 + t_2 D_2 + \sum_{a=3}^{h^{1,1}} t_a D_a \:,
\end{equation}
where we use the same symbols for the divisors and the Poincar\'e dual two forms. We will furthermore restricts to CYs for which it is be possible to find a basis of integral four-cycles such that $D_a$ are orthogonal to both $D_1$ and $D_2$ (this will in general not be an integral basis for $H^{1,1}(X_3,\mathbb{Z})$). This means that the intersection form will have the form
\begin{equation}\label{I3generic}
 I_3= \kappa_1 D_1^3 + \kappa_2 D_2^3 + \kappa_{abc}D_aD_bD_c\:,
\end{equation}
where $a,b,c \neq 1,2$.

In order for the mechanism to work, we will choose $\cF_D$ such that only $t_1$ and $t_2$ appear in the linear combination $\int_{D_D} \cF_D\wedge J$. In the chosen setup, $t_1$ and $t_2$ measure the sizes of $D_1$ and $D_2$ (which are wrapped by the E3 instantons). 
Under the assumption that the non-zero matter fields VEVs  are such that the first term in the D-term potential \eqref{D-term potential} is zero, then the D-term condition (at leading order in $1/\mathcal{V}$ expansion) becomes
\begin{equation}
 \zeta_D = 0\:.\label{xi0}
\end{equation}
Expanding the divisor $D_D$ and the flux in the following way\footnote{The two-form $\cF_D$ belongs to $H^2(D_D)$. In this paper we will only consider two-forms that are pulled back from two-forms of the Calabi-Yau $X_3$; for this reason we will omit the pull-back symbol $i^\ast$ in the expression for the gauge flux, which otherwise should be written as $\mathcal{F}_D = f_1 i^\ast D_1 + f_2 i^\ast D_2$.} 
\begin{equation}
 D_D= d_1 D_1+d_2 D_2 +\sum_a d_a D_a \qquad \mbox{ and } \qquad  \mathcal{F}_D = f_1 D_1 + f_2 D_2\,,\label{DDFDdef}
\end{equation}
with non-zero flux quanta $f_1, f_2$. Eq.~\eqref{xi0} implies
\begin{equation}
 \kappa_1 d_1 f_1 t_1 = \kappa_2 d_2 f_2 t_2 \qquad \Rightarrow \qquad \tau_2 = \frac{\kappa_1}{\kappa_2} \left(\frac{d_1 f_1}{d_2 f_2} \right)^2 \tau_1 \equiv c\, \tau_1\,,\label{t1t2prop}
\end{equation}
where we used $\tau_i = \re T_i=\frac12 \int_{D_i} J \wedge J$ and $J= t_1 D_1 + t_2 D_2 + t_a D_a$. It is essential for \eqref{t1t2prop} to occur that $D_D$ non-trivially intersects  $D_1$ {\it and} $D_2$. The important constraint is that $\cF_D$ has zero coefficients with respect to the $D_a$'s: this is not automatically possible, as Freed-Witten flux quantisation may imply to have these components different from zero.
Only if both conditions are fulfilled then $\zeta_D = 0$ constrains the two volumes $\tau_1$ and $\tau_2$ to be proportional to each other. This relation is what allows the uplift mechanism discussed in the next section.

\subsection{F-term potential\label{sec:f-term-uplift}}

We present a simplified model with $h^{1,1}=3$, constructed by compactifying Type IIB string theory on a Calabi-Yau three-fold with three K\"ahler moduli.
The K\"ahler potential of the K\"ahler moduli space is (see also Section \ref{Sec:ModStabIIB}):
\begin{align}\label{Kthreemod}
 \begin{aligned}
  K= - 2 \ln \left({\cal V} + {\hat{\xi} \over 2}\right) \:.
\end{aligned}
\end{align}
If the intersection form is (\ref{I3generic}), the volume of the three-moduli Calabi-Yau can be written in the Swiss-Cheese form:
\begin{align}
\begin{aligned}
  {\cal V} &= \frac16\left( \kappa_b t_b^3 + \kappa_1 t_1^3 + \kappa_2 t_2^3 \right)
  =\gamma_b (T_b + \bar{T}_b)^{3/2} - \gamma_1 (T_1 + \bar{T_1})^{3/2} - \gamma_2 (T_2 + \bar{T_2})^{3/2}\:,
 \end{aligned}
\end{align}
with one big four-cycle $D_b$ with volume $\re T_b=\tau_b$ and two small cycles with volumes $\re T_{1,2}=\tau_{1,2}$. We used the relation between $t_i$ and $\tau_i$:
\begin{equation}
 t_i = \pm \sqrt{\frac{2 \tau_i}{\kappa_i}}\qquad \text{with} \qquad \gamma_i = \frac{1}{6 \sqrt{\kappa_i}}\,.\label{24cyclerelation}
\end{equation}
The K\"ahler cone condition unambiguously tells us which sign to use in the first equation of~\eqref{24cyclerelation}. In particular, for one big cycle $t_b$ and $h^{1,1}-1$ small cycles $t_i$ one has $t_b > 0$ and $t_i < 0$.

In the following we set $\gamma_b = \gamma_1 = \gamma_2 = 1$ for simplicity. The F-term potential given by \eqref{Kthreemod} and~\eqref{Wthreemod} is the $\mathcal{O}(\mathcal{V}^{-3})$ subleading F-term potential \eqref{SubLeadFtermPot}.
In the model we are presenting here, it is given by\footnote{$\im T_b$ will eventually be stabilized by non-perturbative effects that are omitted in (\ref{Kthreemod}) since they are exponentially suppressed by the CY volume and hence negligible.}
\begin{equation}
 \begin{split}
  V \sim {3 W_0^2 \hat{\xi} \over 4 {\cal V}^3} + {4 W_0 \over {\cal V}^2} \left(\sum_{i=1}^{2} a_i A_i \tau_i e^{-a_i \tau_i} \cos (a_i \theta_i) \right)  + {2 \sqrt{2} \over 3 {\cal V}} \left(\sum_{i=1}^{2} {a_i^2 A_i^2 \tau_i^{1/2}} e^{-2 a_i \tau_i}\right) \:.
 \end{split}
 \label{effective potential at larger volume 0}
\end{equation}
This potential includes the solutions of the Large Volume Scenario (LVS) \cite{Balasubramanian:2005zx}, where ${\cal V} \sim e^{a_i \tau_i}$, realizing an exponentially large volume.

At leading order the D-term potential stabilizes $\tau_2 = c\,\tau_1$, so that the resulting potential is a function of $\tau_s \equiv (\tau_1 + \tau_2/c)/2 = \tau_1$, while the orthogonal direction $\tau_z =(\tau_1 - \tau_2/c)/2$ is heavy and can be integrated out.
For simplicity of presentation, we make the redefinitions:
\begin{equation}
 \begin{split}
  &x_s = a_1 \tau_s,\quad y_s = a_1 \theta_s = {a_1 \over 2} \left(\theta_1 + {\theta_2 \over c} \right) \quad {\cal V}_x = {\cal V} a_1^{3/2},
  c_i = {A_i \over W_0},\quad  \hat{\xi}_x = \hat{\xi} a_1^{3/2},  
 \end{split}
 \label{fields redefinition}
\end{equation}
and
\begin{equation}
\beta = c \frac{a_2}{a_1}.
\end{equation}
Then the effective potential at order ${\cal O}({\cal V}^{-3})$ becomes\footnote{The imaginary mode of $Z = \tau_z + i \theta_z$ is eaten by the $U(1)$ gauge boson that becomes massive through the St\"uckelberg mechanism. Hence it is integrated out together with its partner $\tau_z$.
}
\begin{equation}
 \begin{split}
  \hat{V} \equiv \left({a_1^3 \over W_0^2}\right) V \sim& {3\hat{\xi}_x \over 4 {\cal V}_x^3}  + {4 c_1 x_s \over {\cal V}_x^2}e^{-x_s} \cos y_s + {2\sqrt{2} c_1^2 x_s^{1/2}\over 3 {\cal V}_x} e^{-2 x_s }\\
  &+ {4 \beta c_2 x_s \over {\cal V}_x^2} e^{-\beta x_s} \cos (\beta y_s) + {2\sqrt{2} \beta^2 c_2^2 x_s^{1/2}\over 3 {\cal V}_x} e^{-2 \beta x_s } .
 \end{split}
 \label{redefined potential}
\end{equation}

The potential (\ref{redefined potential}) is the starting point for the uplift mechanism for dS vacua that has been proposed in \cite{Rummel:2014raa}.
Roughly speaking, the first line of the potential (\ref{redefined potential}) is nothing but the potential for LVS (that would produce and $AdS$ minimum),  while the second line plays a role of the uplifting term. In practice, one stabilizes the potential \eqref{redefined potential} and finds a dS minimum by gradually increasing $c_2 >0$, while keeping $c_1<0$. 

We obtain a Minkowski vacuum when $\partial_{\Vol_x, x_s} \hat{V} =0, \, \hat{V}=0$, which happens for 
\begin{equation}
 \begin{split}
  \hat{\xi}_x \sim 4\sqrt{2} x_s^2, \quad c_1 \sim - {3 \sqrt{x_s} \over \sqrt{2} \Vol_x} e^{x_s}, \quad c_2 \sim {9 \over 4\sqrt{2x_s} \Vol_x \beta(1- \beta)} e^{\beta x_s},
 \end{split}
\end{equation}
where we have used an approximation of large $x_s$ and small $c_2$, and $y_s$ is stabilized at zero.
Plugging the conditions in the Hessian, we obtain
\begin{equation}
 \begin{split}
  \partial_{\Vol_x}^2 \hat{V}|_{\rm ext} \sim {6\sqrt{2} x_s^{3/2} \over \Vol_x^5}, \quad \det\left(\partial_{i} \partial_j \hat{V} \right)|_{\rm ext} \sim {54 (1-\beta) x_s^2 \over \Vol_x^8}, \quad \partial_{y_s}^2 \hat{V}|_{\rm ext} \sim {6\sqrt{2} x_s^{3/2} \over \Vol_x^3},
 \end{split}
\end{equation}
where $i,j=\Vol_x, x_s$ and $\partial_i \partial_{y_s} \hat{V}|_{\rm ext} =0$.
According to Sylvester's criterion, the stability is ensured when these three quantities are defined positively, suggesting $\beta<1$.
Increasing (slightly) the values of $c_2$, one obtains a (tiny) dS minimum.

This mechanism does not work if we start with a CY with two K\"ahler moduli $\tau_b$ and $\tau_s$ and a racetrack superpotential 
$W = W_0 + A_1 e^{-a_1 T_s} + A_2 e^{-a_2 T_s}$. In this case there would appear a cross term proportional to $c_1, c_2$ in the potential. Repeating the same analysis as above and requiring a Minkowski minimum, one runs into a contradiction. This is related to the fact that the cross-term contributes negatively to the potential, disturbing the uplift: making the uplift term larger to overcome this negativity destabilizes the vacuum.
Due to the absence of this dangerous cross-term, the uplift mechanism works well in the D-term generated racetrack model.

\subsection{Summary of conditions for de Sitter vacua}

Let us summarize what are the conditions for the dS uplift to work.
\begin{itemize}
\item We need a Calabi-Yau three-fold with two rigid divisors $D_1$ and $D_2$ (with $h^{1,0}=0$)  that support non-perturbative effects generating the non-perturbative superpotential \eqref{Wthreemod}. We restrict to the case when these two divisors do not intersect each other. In particular we also require them to be part of a basis of divisors with intersection form \eqref{I3generic}.
\item The second ingredient is the presence of D7-branes with a non-trivial gauge flux. We require the D7-brane divisor $D_D$ to intersect the two rigid divisors $D_1$ and $D_2$. The flux $\cF_D$ will generate an FI-term in the D-term potential that depends on the volumes of the two rigid divisors. Under the assumption that the contribution to the D-term from the open string moduli is zero, the D-term condition boils down into a linear relation between the two volumes.
\item The flux $\cF_D$ should be chosen not to be too large, in order to still satisfy the D3-tadpole cancellation condition, and such that 
 \begin{equation}
  \beta \equiv c \frac{a_2}{a_1} \neq 1\,,
 \end{equation}
where $c$ depends on the flux quanta and the divisor choice, see \eqref{t1t2prop}.
The uplift term is identified to be the term of $e^{-a_2 t_2}$ when $\beta < 1$, while $e^{-a_1 t_1}$ for $\beta > 1$.
When the two non-perturbative terms in the superpotential are generated by E3 instantons, we have $a_i = 2\pi$. 
\end{itemize}
If $c =1$, we may still get de Sitter uplift if at least one of the non-perturbative effects is generated by D7-brane gaugino condensation for which $a_i = 2\pi/N_i$ where $N_i$ is the Coxeter number of the corresponding gauge group.
It is also possible to realize the scenario in a more general setup, but this gives rise to further constraints. 
For instance, if the two small cycles intersect, their intersection numbers have to fulfil certain requirements for non-vanishing non-perturbative effects (see Appendix of~\cite{Rummel:2014raa}). One could also consider non-rigid divisors that can generate a non-zero non-perturbative superpotential if they support a proper flux that lifts the exceeding zero modes  \cite{Bianchi:2011qh,Bianchi:2012kt}.

\section{Searching for toric divisors}\label{Sec:SearchToric}

In this section we explain how to efficiently find Calabi-Yau threefolds $X_3$ for which there exist two divisors $D_1$ and $D_2$ which are rigid (i.e.
$h^{2,0}(D_1) = h^{2,0}(D_2) = 0$), irreducible and for which the intersection form can be written as
\begin{equation}\label{intformnice}
\kappa_{1} D_1^3 + \kappa_{2} D_2^3 + \sum_{a,b,c \neq 1,2} \kappa_{abc} D_a D_b D_c \, ,
\end{equation}
for a basis of $H^{1,1}(X_3)$ formed by $D_1$, $D_2$ and $\{D_a\}$. Furthermore, we are going to demand that the K\"ahler cone is such that
\eqref{24cyclerelation} can be fulfilled, i.e. $D_1$ and $D_2$ are two `small' cycles.

In order to have a large set of examples while keeping technical control we restrict ourselves to Calabi-Yau threefolds which are hypersurfaces
in toric varieties and divisors $D_1$, $D_2$ which descend from toric divisors. A toric hypersurface Calabi-Yau threefold is constructed from a pair or four-dimensional reflexive polytopes $\Delta^\circ$ and $\Delta$ as well as an appropriate triangulation of $\Delta^\circ$. A polytope $\Delta \subset \mathbb{R}^4$ with vertices on a lattice (which we can take to simply be $\mathbb{Z}^4 \subset \mathbb{R}^4$) is called reflexive if its polar dual, defined by
\begin{equation}
\langle \Delta, \Delta^\circ \rangle \geq -1 \:,
\end{equation}
is a lattice polytope as well. In this case there is a one-to-one relationship between $k$-dimensional faces $\Theta^{\circ [k]}$ of $\Delta^\circ$
and $(3-k)$-dimensional faces of $\Theta^{[3-k]}$ of $\Delta$.
This data defines the Calabi-Yau as follows: from the triangulation, one can construct a fan $\Sigma$ over the faces of the polytope $\Delta^\circ$ which gives rise to a four-dimensional toric variety $\PP^4_{\Sigma}$. The lattice points on $\Delta$ determine the complete linear system of $-K_{\PP^4_{\Sigma}}$
and hence a Calabi-Yau threefold hypersurface $X_3$.  For Calabi-Yau threefolds, i.e. four-dimensional polytopes, there always exists a 
triangulation resulting in a smooth Calabi-Yau threefold $X_3$ \cite{Batyrev:1994hm}. Crude topological invariants such as the Hodge
numbers only depend on the dual pair of polytopes and not on the details of the triangulation. In particular,
\begin{equation}\label{eq:hodgecomb}
\begin{aligned}
h^{1,1}(X_3) &= \ell(\Delta^\circ) - 5 - \sum_{\Theta^{\circ[3]}} \ell^*(\Theta^{\circ[3]}) + \sum_{(\Theta^{\circ[2]},\Theta^{[1]})}  \ell^*(\Theta^{\circ[2]})\ell^*(\Theta^{[1]}) \\
h^{2,1}(X_3) &= \ell(\Delta) - 5 - \sum_{\Theta^{[3]}} \ell^*(\Theta^{[3]}) + \sum_{(\Theta^{[2]},\Theta^{\circ[1]})}  \ell^*(\Theta^{[2]})\ell^*(\Theta^{\circ[1]}) \, .
\end{aligned}
\end{equation}
Here, $\ell^*$ counts interior lattice points of a face and $\ell(\Delta)$ counts all lattice points on the polytope $\Delta$.\footnote{In this construction, mirror symmetry is realized by swapping the roles played by the pair of dual polytopes $\Delta \leftrightarrow \Delta^\circ$.}

The more refined data needed here, such as the intersection form and the K\"ahler cone, is not determined by the polytopes alone, but depends on the triangulation. While all four-dimensional reflexive polytopes have been famously classified in \cite{Kreuzer:2000xy}, all possible triangulations
are not known for each polytope. Recently, \cite{Altman:2014bfa} have determined all triangulations for all reflexive polytopes which give
rise to Calabi-Yau threefolds with $h^{1,1}(X_3) \leq 6$. Using their results, it is possible to completely answer for which toric hypersurface
Calabi-Yau threefolds (of $h^{1,1}(X_3) \leq 6$) the intersection form can be written as \eqref{intformnice}, if we restrict to the cases where the two small 
divisors $D_1$ and $D_2$ of $X_3$ descend from toric divisors of the ambient space $\PP^4_{\Sigma}$.

\subsection{Hodge numbers and geometry of toric divisors}

The reason we have singled out toric divisors is that they are both easy to enumerate and to analyse. The fan $\Sigma$ of the ambient toric variety
containing $X_3$ is constructed from a triangulation of $\Delta^\circ$, so that every lattice point $\nu_i$ on $\Delta^\circ$ generates a ray in $\Sigma(1)$ and hence corresponds to a homogeneous coordinate $z_i$ and a toric divisor $T_i$ of $\PP^4_\Sigma$. Toric divisors on $\PP^4_\Sigma$ give rise to
divisors on $X_3$ by restriction, or equivalently, by the pullback $i^*$ associated with the embedding $i: X_3 \mapsto \PP^4_{\Sigma}$.
The Hodge numbers $h^{p,0}$ of a divisor $D_i = i^* T_i$ only depend only on the location of the point $\nu_i$ in the polytope $\Delta^\circ$ \cite{Danilov1987,Batyrev1993}:
\begin{itemize}
 \item If $\nu_i$ is a vertex of $\Delta^*$,  $D_i$ is always an irreducible divisor on $X_3$. The hodge number $h^{2,0}(D_i)$
 is given by $\ell^*(\Theta^{[3]})$, where $\Theta^{[3]}$ is the dual face (on $\Delta$) to $\nu_i = \Theta^{\circ[0]}$.  The hodge number
 $h^{1,0}(D_i)$ always vanishes in this case.
 \item If $\nu_i$ is inside a one-dimensional face $\Theta^{\circ[1]}$ of $\Delta^\circ$, $D_i$ is always irreducible and $h^{2,0}(D_i)=0$. The Hodge number
 $h^{1,0}(D_i)$ is given by $\ell^*(\Theta^{[2]})$, where  $\Theta^{[2]}$ is the dual face to  $\Theta^{\circ[1]}$.
 \item If $\nu_i$ is inside a two-dimensional face $\Theta^{\circ[2]}$ of $\Delta^\circ$, $D_i$ is always rigid, $h^{2,0}(D_i)=0$, and has 
 $h^{1,0}(D_i)=0$, but it is reducible in general. Its number of irreducible components is given by $\ell^*(\Theta^{[1]})$, where $\Theta^{[1]}$ is the dual face to
 $\Theta^{\circ[2]}$. For this reason, the divisors $\{D_i\}$ do not generate all of $h^{1,1}(X_3)$ and the combinatorial equation \eqref{eq:hodgecomb}
 contains the `correction term' 
 \begin{equation*}
  \sum_{(\Theta^{\circ[2]},\Theta^{[1]})}  \ell^*(\Theta^{\circ[2]})\ell^*(\Theta^{[1]}) \, .
 \end{equation*}
 \item If $\nu_i$ is inside a three-dimensional face of $\Delta^\circ$, $T_i$ does not intersect the Calabi-Yau hypersurface at all, i.e. 
 there is no associated $D_i$. This is the reason for the subtraction of 
 \begin{equation*}
   \sum_{\Theta^{\circ[3]}} \ell^*(\Theta^{\circ[3]})\,,
 \end{equation*}
in \eqref{eq:hodgecomb}. Furthermore, such points can be omitted in a triangulation of $\Delta^\circ$, as done in \cite{Altman:2014bfa}.
\end{itemize}
The remaining Hodge number $h^{1,1}$ depends on the triangulation chosen.

Hence we can summarize the condition for a toric divisor to restrict to a divisor $D$ on $X_3$ which is irreducible and has $h^{1,0}(D_i)=h^{2,0}(D_i)=0$ by\footnote{Finally, let us comment on the condition for rigidity of divisors inside a Calabi-Yau threefold. If a
divisors $D_i$ is such that $h^{2,0}(D_i) = 0$, there can be no global sections in its canonical bundle. By adjunction, and because $X_3$ is a Calabi-Yau manifold, we have the relation $
 K_D = N_{D \subset X_3}
$ between bundles on $D$. Hence there can also not be any global holomorphic section of the normal bundle of $D$ inside $X_3$, i.e. the divisor is rigid.} 
\begin{equation}\label{eq:condirredrigid}
 \ell^*(\Theta) = 0 \, ,
\end{equation}
where $\Theta$ is the dual to the face containing $\nu_i$. We only need to consider lattice points on the 2-skeleton of $\Delta^\circ$, as points interior to 
three-dimensional faces of $\Delta^\circ$ give rise to toric divisors which do not meet $X_3$.

\subsection{Intersection ring}

The intersection ring (Chow ring) of the ambient space $\PP^4_{\Sigma}$ can be easily read off from the fan (or from the triangulation of $\Delta^\circ$).
two divisors (or, more generally, codimension $p$ algebraic subvarieties) intersect iff they share a common cone in $\Sigma$. More concretely, four divisors $T_i$ intersect in a point iff they span a cone of volume one\footnote{If the cone has volume $A$, the intersection number is $1/A$.} in $\Sigma$.

As we have discussed already, the second cohomology of the Calabi-Yau threefold is not generated by the $T_i$ alone, as some of these divisors $D_i$ are reducible and the components give rise to a different classes in $H^{1,1}(X_3)$ in general. Performing computations on the ambient space $\PP^4_{\Sigma}$ 
hence only gives us access to part of the intersection ring of the Calabi-Yau hypersurface $X_3$. 
The cup product between a cycle $D_i$ descending from the ambient space (in the image of $i^*$) and a cycle $D_a$ intrinsic to $X_3$ (in the cokernel of $i^*$) is \cite{Dodson1997}
\begin{equation}\label{eq:cokimorth}
 D_i \cdot D_a = i^*T_i \cdot D_a = i^*PD( T_i \cdot i_* D_a) = 0 \, .
\end{equation}
More explicitly, this can described as follows: if a toric divisor $T_i$ becomes reducible on the Calabi-Yau threefold $X_3$ we can write
\begin{equation}
 D_i = \sum_\mu D_i^\mu \, ,
\end{equation}
where $D_i^\mu$ are its irreducible components. Only the sum above lifts to a divisor of the ambient space and the linear combinations
\begin{equation}\label{eq:cokeri*}
\{ D_i^\mu - D_i^\nu \, \,\,\forall i,\mu,\nu \}\, ,
\end{equation}
which generate the cokernel of $i^*$, are orthogonal to all divisors $D_k$ descending from $\PP^4_\Sigma$.

Hence the intersection form will split into a piece descending from the toric ambient space and an intrinsic piece generated by \eqref{eq:cokeri*}.
Asking weather we can achieve the form \eqref{intformnice} for two given toric divisors $D_1$ and $D_2$, 
we can hence safely focus on divisors $D_i$ descending from the ambient space.

\subsection{Orthogonality}

We can now address the central problem and ask when, for two irreducible and rigid divisors $D_1 = i^* T_1$ and $D_2 = i^* T_2$, 
we can achieve the form \eqref{intformnice}. As argued in the section above, we can restrict $H^{1,1}(X_3)$ to
divisors descending from the ambient space $\PP^4_\Sigma$. We denote the subspace of $H^{1,1}(X_3)$ in the image
of $i^*$ by $H^{1,1}_T(X_3)$.

In order to simplify things, we first show that \eqref{intformnice} is equivalent to the statement that 
\begin{align}\label{eq:orthocond}
 D_1 \cdot D_i = 0 \qquad\forall i \neq 1 \qquad\mbox{and}\qquad
 D_2 \cdot D_i = 0 \qquad\forall i \neq 2 \,,
\end{align}
as an equation in (co)homology. Clearly \eqref{intformnice} follows from this, so the nontrivial statement is the
converse. To see this, let us assume that \eqref{intformnice} holds for a basis of $H^{1,1}(X_3)$ and
there is an $i$ such that $D_1 \cdot D_i   \neq 0$ in cohomology. By Poincar\'e duality, there must then be a cycle $D_{1i}$ such that
\begin{equation}
 D_{1i}\cdot D_1 \cdot D_i = 1 \, .
\end{equation}
As we have started out with a basis, we can also expand $D_{1i} = \sum_k a_ k D_k$. The above now says that there must be an $k$ such that
$t_{11k}$ or $t_{1ki}$ is non-zero (or both). This is a contradiction, and we are done after repeating the same argument for $D_2$. 

The above greatly simplifies matters, as it gives us a way to show if we can reach a basis in which \eqref{intformnice} holds
in a constructive fashion. We simply have to find enough linear combinations of toric divisors orthogonal to both $D_1$
and $D_2$ to form a basis of $H^{1,1}_T(X_3)$. If this can be done, we have reached a basis realizing \eqref{intformnice}.
If this is not possible, then there is no such basis for the specific choice of $D_1$ and $D_2$ made.

\subsection{Computations in practice}

The above discussion can be straightforwardly cast into a practical algorithm to find all solutions to \eqref{intformnice} for pairs of toric divisors $D_1$ and $D_2$ and toric Calabi-Yau hypersurfaces $X_3$. The necessary computations can be conveniently carried out using
the routines dealing with polytopes and toric geometry included in the ever-helpful sage \cite{Stein2014}.

As our starting point, we assume that we are given a dual pair of reflexive
polytopes $\Delta$ and $\Delta^\circ$ as well as a triangulation of $\Delta^\circ$. In practice, we taken this data from the tables
available online described in \cite{Altman:2014bfa}. 

\begin{itemize}
 \item From the list of toric divisors $\{T_i\}\simeq \Sigma(1)$, determine the set $\{T_{IR}\}$ divisors which are irreducible and have $h^{1,0}(D_i)=h^{2,0}(D_i)=0$ by
 using \eqref{eq:condirredrigid}.
 \item For each pair of toric divisors $D_1$ and $D_2$ in $\{T_{IR}\}$
 \begin{itemize}
 \item Check $D_1 \cdot D_2 = 0$.
 \item Check if we can find enough linear combinations orthogonal to $D_1$ and $D_2$ such that we can form a basis of $H^{1,1}(X_3,\mathbb{R})$. If this is possible, we have reached the form
 \eqref{intformnice}.
 \item Check that the K\"ahler cone is such that $D_1$ and $D_2$ are small, that for the chosen intersection form means $\kappa_1 t_1 < 0 $ and $\kappa_2 t_2 < 0 $.
 \item If all these checks are successful, we can compile a list of toric divisors $D_D$ which are irreducible and intersect both $D_1$ and $D_2$ in order to be wrapped by a magnetized $D7$-brane.
 Again, these are not the only interesting candidates for $D_D$, but can be most easily described. 
 \end{itemize}
 
\end{itemize}

\section{Uplift mechanism for Calabi-Yau hypersurfaces in toric varieties: scan over geometries and fluxes}\label{Sec:SearchFlux}

In the previous section we have outlined how to find Calabi-Yau manifolds with a topology suitable for the wanted dS uplift mechanism for the case of hypersurfaces
in toric varieties. Using the database of triangulations of polytopes leading to Calabi-Yau manifolds with small $h^{1,1}$ worked out by \cite{Altman:2014bfa}, 
we made a scan for $h^{1,1}(CY)$ up to 4, and counted the manifolds that allow our geometric conditions to be satisfied.
In particular,  we first consider all Calabi-Yau manifolds that allow an intersection form like \eqref{I3generic} with $D_1$ and $D_2$ two rigid divisors. We then look for (irreducible) toric divisors $D^{\rm toric}_i=\{z_i=0\}$ that intersect the two rigid divisors $D_1$ and $D_2$. 
For each of them we construct a different model, with the D7-brane divisor $D_D$ wrapping it a number of times. For the ones with larger Euler characteristic the negative D3-charge is maximized (see \eqref{QD3D7}), allowing a bigger choice of fluxes that saturate the D3-tadpole cancellation condition. 
We find that the geometric criteria for realizing the D-term generated racetrack scenario are fulfilled by roughly a subset of $15 \%$ in the case of $h^{1,1}=3$ and $24 \%$ in the case of $h^{1,1}=4$.

The orientifold involution will be chosen such that the fixed point divisor $D_\xi=\{\xi=0\}$ is proportional to $D_D$. The D7-tadpole will be cancelled by taking either four D7-branes (plus their for image branes) on top of the O7-locus, or one D7-brane (plus its image brane) wrapping divisors in the class $D_D=4D_\xi$.\footnote{In this way, we avoid the presence of several D7-brane stacks, that would be difficult to control. They will typically intersect the divisors $D_1$, $D_2$ and $D_D$ and spoil the mechanism.}
Furthermore, we can have the leading non-perturbative contribution to the superpotential to be rank-one ED3-instantons or rank-two instantons (by forbidding rank-one instantons). This gives us a total of four possible setups that we discuss separately in the following.

\subsection{Four D7 branes and rank-one instantons}\label{fourD7oneinst_sec}

By taking four D7-branes (plus their four images) on top of the O7-plane locus, we get an D7-brane stack supporting the gauge group $SO(8)$ on the divisor $D_D=D_\xi$. In order to realize explicit models of the D-term generated racetrack, we need to specify the gauge flux $\mathcal{F}_D$ living on the worldvolume of the D7-branes. We take the same flux on all the four D7-branes (and $-\cF_D$ on their orientifold images). The tadpole cancellation condition can then be saturated by three-form fluxes and D3-branes if $Q_{D3}^{O7+4D7}-\frac12 n_{O3}<0$, where $Q_{D3}^{O7+4D7}$ is given in \eqref{O74D7QD3} (with $D_\xi=D_D$ and $\cF=\cF_D$). Using the assumed form of $D_D$ and $\cF_D$ in \eqref{DDFDdef}, we have the condition:
\begin{equation}
  \frac{\chi(D_D)}{2} + 4 \left(d_1 \kappa_1 f_1^2 + d_2 \kappa_2 f_2^2 \right) > 0 \,,\label{D3Dracetrack}
\end{equation}
where the first contribution is positive, while the second is negative definite (when $\zeta_D=0$).
We see that choosing $D_D$ with large $\chi(D_D)$ is beneficial in terms of a large flux choice $f_1$ and $f_2$ as well as possible $F_3$ and $H_3$ fluxes which are necessary for complex structure moduli stabilization. For the models that pass the criteria outlined in the previous section, we find either two or three possible choices for $D_D$.

The gauge flux is given as $\mathcal{F}_D = F_D - i^\ast B$. $F_D$ must be quantized according to \eqref{FWanomalyCancel}. Let us consider an integral basis of $H^{1,1}(X_3,\mathbb{Z})$, $\{D^{\text{int}}_i\}$ with $i=1,...,h^{1,1}(X_3)$\footnote{All elements in the integral cohomology are linear combination of the integral basis elements with integral coefficients.}. 
Hence the gauge flux 
\begin{equation}
 F_D = \sum_{i=1}^{h^{1,1}} F_i D^{\text{int}}_i + \frac{c_1(D)}{2}\,, \qquad\qquad \mbox{with}\qquad F_i\in\mathbb{Z}  \,,
\end{equation}
is properly quantized. 
On the E3 instantons wrapping the divisors $D_1$ and $D_2$ we choose the flux to be $F_i = -c_1(D_i)/2 = D_i/2$ . 
The globally defined background B-field can be also expanded in the integral basis with integral or half-integral coefficients:
\begin{equation}
 B =  \sum_{i =1}^{h^{1,1}} B_i D^{\text{int}}_i\,,  \qquad \qquad \mbox{with} \qquad B_i \in \mathbb{Z}/2 \:.
\end{equation}

In order to have invariant rank-one instantons (we are assuming that the $D_i$ are odd cycles), the $B$-field must make $\cF_1$ and $\cF_2$ vanish. This happens in our setup when
\begin{equation}
 B = \frac{D_1}{2} + \frac{D_2}{2} + \sum_{a \neq 1,2} B_a D_a\,,
\end{equation}
where we used the basis with $D_1$, $D_2$ and orthogonal divisors $D_a$. The coefficients $B_a$ must be chosen such that $B$ cancels possible coefficients of $F_D$ along $D_a$ that may be enforced by Freed-Witten anomaly cancellation. Only if this is possible, we will have $\cF_D=f_1D_1+f_2D_2$.
This condition is given in term of the diagonal basis $\{D_1,D_2,D_a \}$, which does not coincide generically with the integral basis $\{D^{\text{int}}_i \}$. Hence, the $B_a$ are not necessarily in $\mathbb{Z}/2$. 
Luckily, in our setup (hypersurfaces in toric ambient spaces) the relation between the two bases is easy to find.
Using the basis transformation, we can then determine what are the allowed values of $f_1$ and $f_2$ and if it is possible to set to zero all the other coefficients $f_a$ for $a>2$.

\begin{figure}[t!]
\centering
\includegraphics[width= 0.47\linewidth]{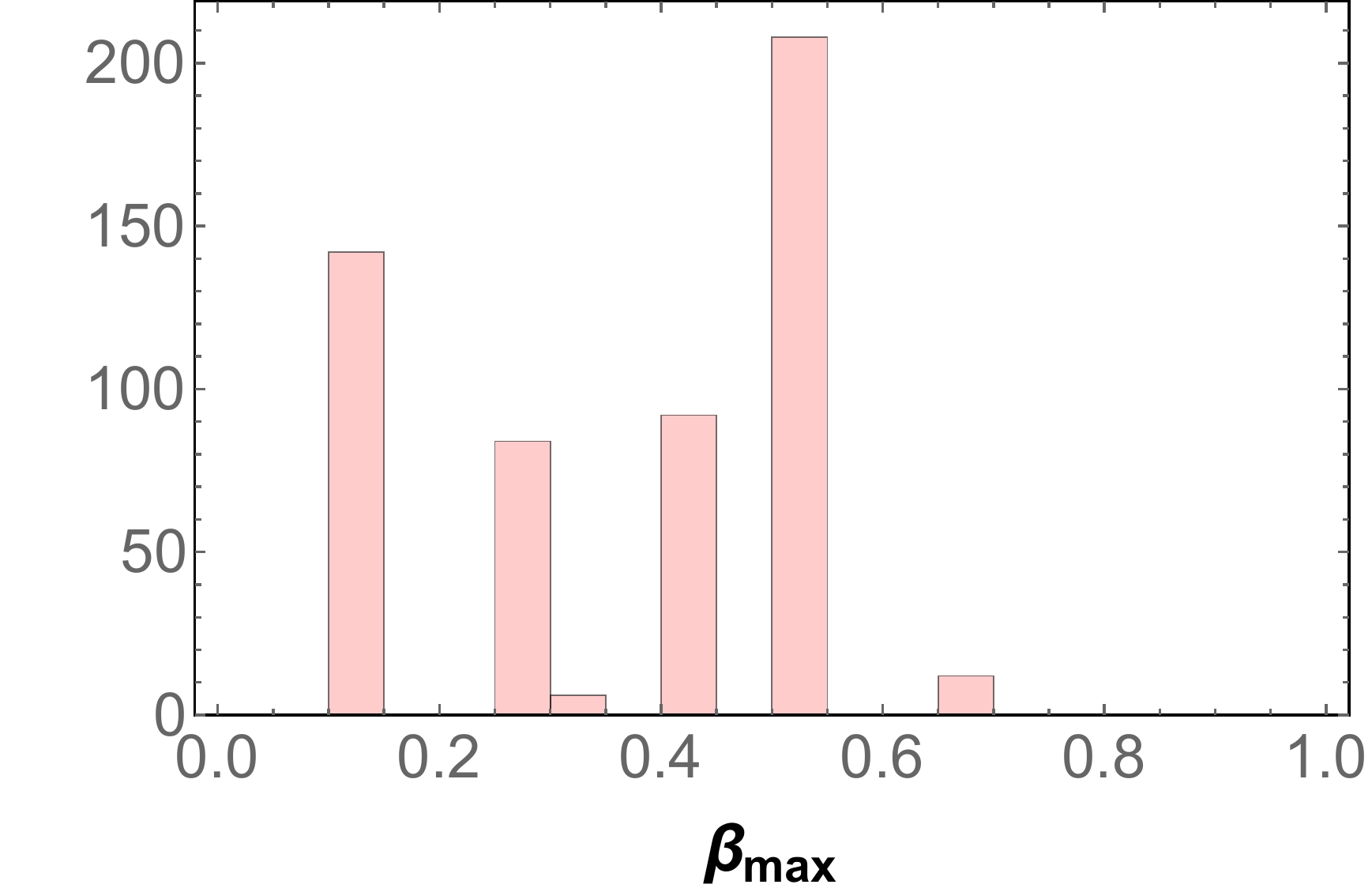}
\includegraphics[width= 0.48\linewidth]{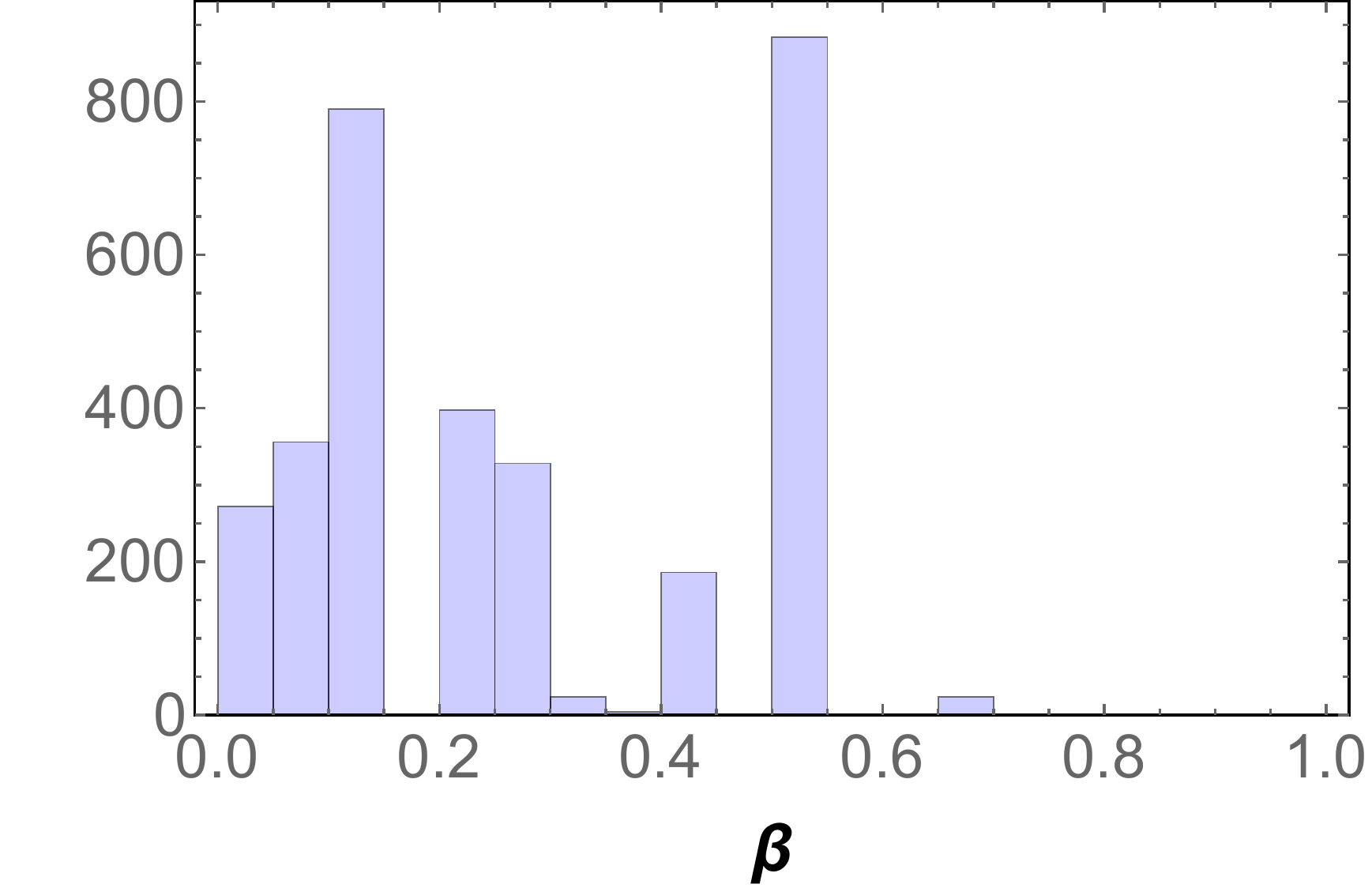}
\caption{Four D7 branes and rank-one instanton: on the left, we show the distribution of $\beta_{\text{max}}$, i.e. the maximal value of $\beta$ for each polytope that we find in our scan. On the right we show all possible values of $\beta$ and their relative distribution in our scan.}
\label{betadistr_fig1}
\end{figure}

We supplement the scan over CY manifolds by a scan over all possible fluxes $F_i$ and B-field coefficients $B_i$ that fulfil the conditions outlined above.
The results are summarized in Table \ref{sec4res1_tab}. The first column contains the number of polytopes for a given $h^{1,1}$, while the second column contains the number of polytopes where the geometric conditions of the simplest incarnation of the D-term generated racetrack scenario can be met, i.e. two non-intersecting rigid divisors and at least one irreducible divisor that intersects the two. Then, we scan over all possible fluxes allowed by the D3 tadpole \eqref{D3Dracetrack}. The polytopes for which one can find a gauge flux $\mathcal{F}_D$ such that $c \neq 1$ and $I_{D}^{A} \neq 0$ is listed in the third column of Table \ref{sec4res1_tab}. Furthermore, the values of $\beta$ we find in our scan are listed in Figure \ref{betadistr_fig1}. The closest values of $\beta$ that we find in our scan for this setup are $\beta=2/3, 1/2,4/9,9/25,1/3$. For values closer to one we consider different setups in the following.

\begin{table}[ht!]
\centering
  \begin{tabular}{|c|c|c|c|}
  \hline
  $h^{1,1}$ & $\# \text{polytopes}$ & $\# \text{polytopes where geometric}$ & $\# \text{polytopes where geometric conditions are}$\\
   & & $\text{conditions are met}$ & $\text{met and suitable fluxes can be found}$\\
  \hline
  $3$ & 244 & 39 & 10 \\
  \hline
  $4$ & 1197 & 285 & 87\\
  \hline
  \end{tabular}
  \caption{Four D7 branes and rank-one instanton: number of polytopes where the D-term generated racetrack uplift can be applied.}
  \label{sec4res1_tab}
\end{table}

\subsection{One D7 brane and rank-one instantons}\label{oneD7oneinst_sec}

Alternatively, we can take the same orientifold involution as in the previous subsection, but wrap one D7 brane on the divisor $4 D_D$ instead. As a consequence, $F_D$ is an integral form. In practice, this allows to have milder constraints on the flux quanta.
Furthermore, the D3 tadpole cancellation condition changes from \eqref{D3Dracetrack} to
\begin{equation}
  \frac{\chi(D_D)}{6} + \frac{\chi(4 D_D)}{12}+ 4 \left(d_1 \kappa_1 f_1^2 + d_2 \kappa_2 f_2^2 \right) > 0 \,,\label{D3Dracetrack2}
\end{equation}
effectively replacing $\chi(D_D)/2$ by $\chi(D_D)/6 + \chi(4 D_D)/12 \gg \chi(D_D)/2$. Hence, the D3 tadpole condition becomes less restrictive, i.e. more values of $f_1$ and $f_2$ can be considered than in Subsection \ref{fourD7oneinst_sec}.

\begin{figure}[t!]
\centering
\includegraphics[width= 0.47\linewidth]{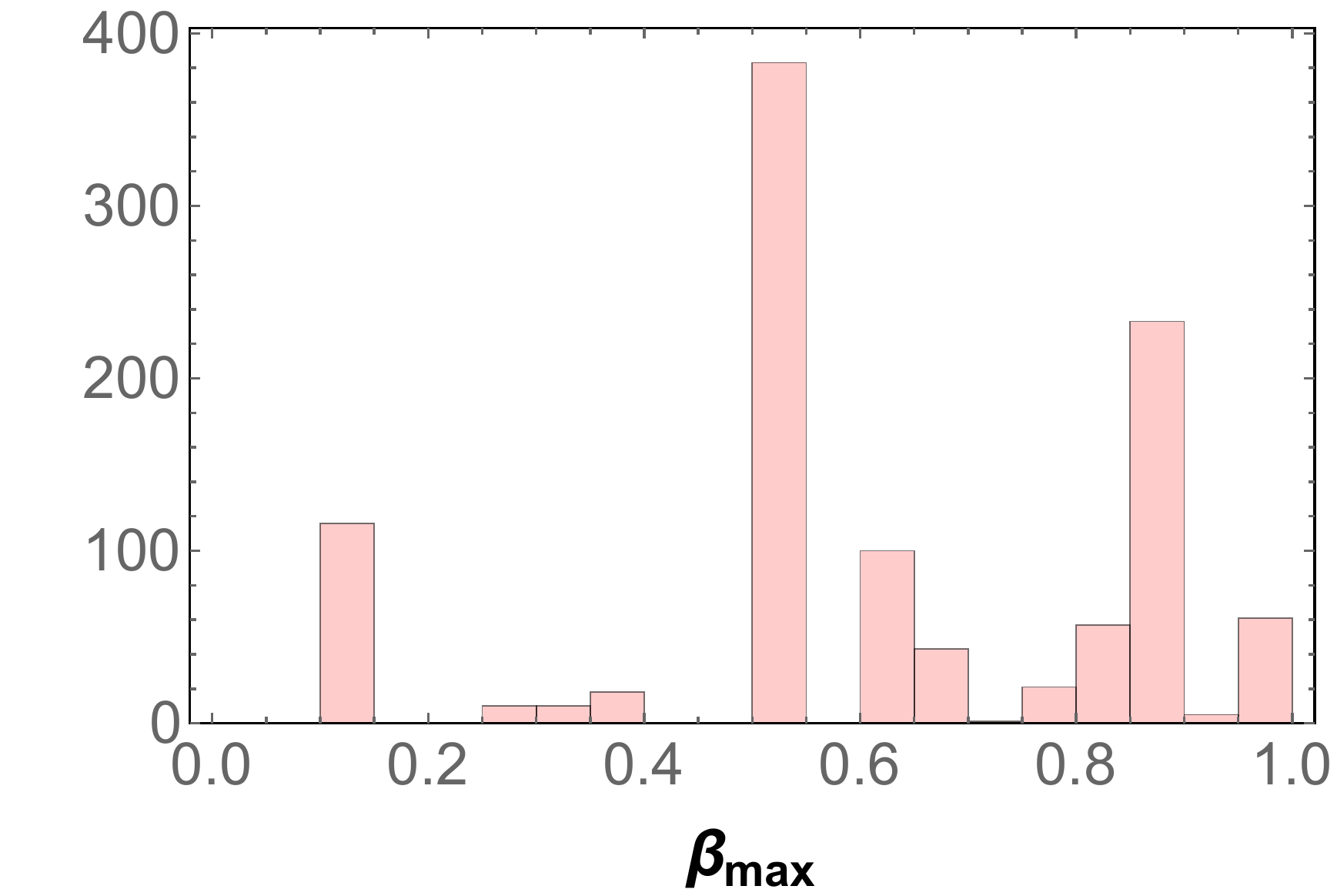}
\includegraphics[width= 0.48\linewidth]{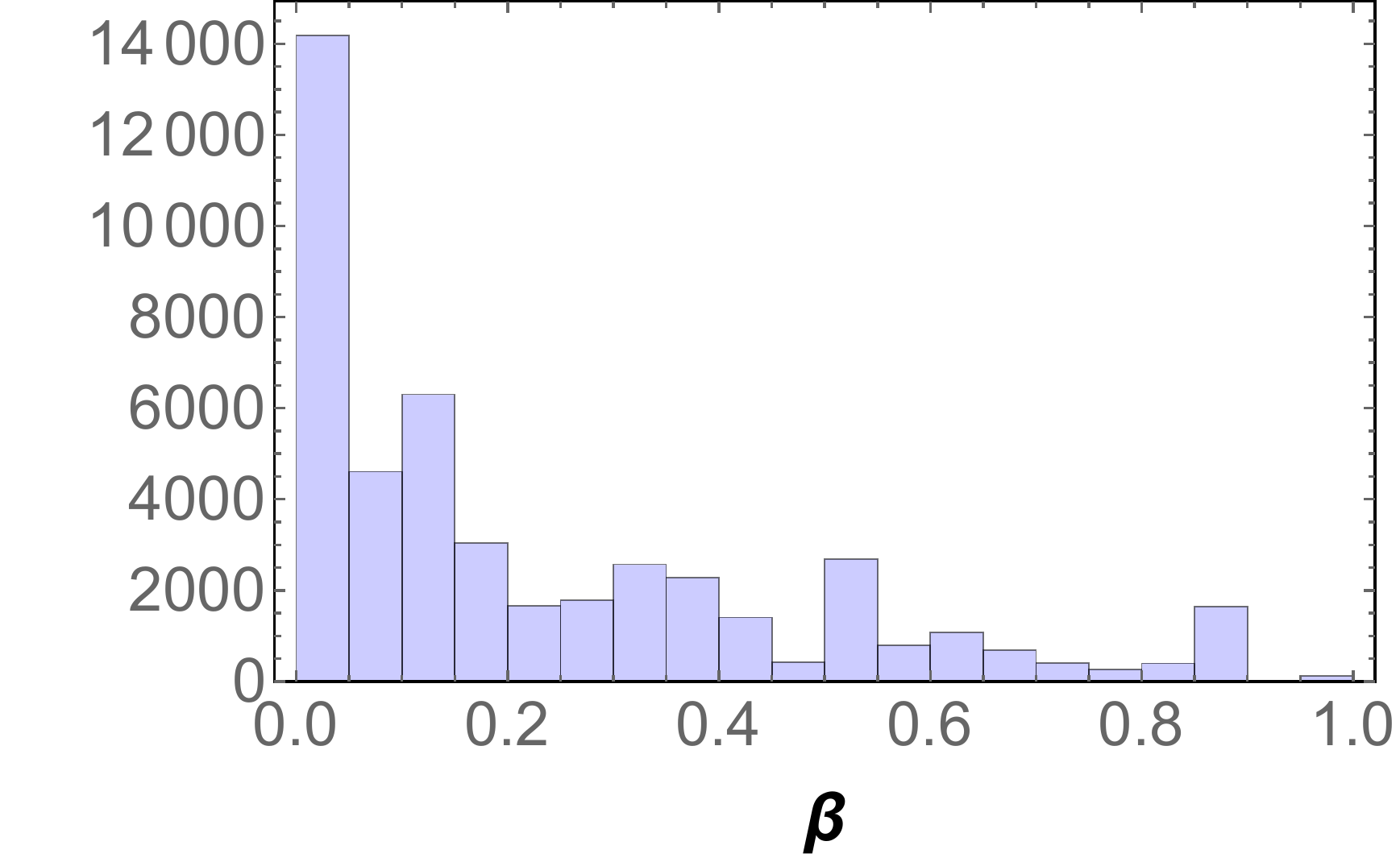}
\caption{One D7 brane and rank-one instanton: on the left, we show the distribution of $\beta_{\text{max}}$, i.e. the maximal value of $\beta$ for each polytope that we find in our scan. On the right we show all possible values of $\beta$ and their relative distribution in our scan.}
\label{betadistr_fig2}
\end{figure}

The results are presented in Table \ref{sec4res2_tab} and Figure \ref{betadistr_fig2}. Clearly a larger fraction of polytopes can accommodate a D-term generated racetrack scenario compared to Section \ref{fourD7oneinst_sec}. Furthermore, there is a large variety in values of $\beta$ and many models with $\beta \lesssim 1$ can be found. The values closest to one we find in our scan are $49/50,121/128,225/242,8/9,169/200$.

\begin{table}[ht!]
\centering
  \begin{tabular}{|c|c|c|c|}
  \hline
  $h^{1,1}$ & $\# \text{polytopes}$ & $\# \text{polytopes where geometric}$ & $\# \text{polytopes where geometric conditions are}$\\
   & & $\text{conditions are met}$ & $\text{met and suitable fluxes can be found}$\\
  \hline
  $3$ & 244 &  39 &  32 \\
  \hline
  $4$ & 1197 &  285 & 191 \\
  \hline
  \end{tabular}
  \caption{One D7 brane and rank-one instanton: number of polytopes where the D-term generated racetrack uplift can be applied.}
  \label{sec4res2_tab}
\end{table}

\subsection{Four D7 brane and rank-two instantons} \label{fourD7twoinst_sec}

Here, we choose the same D7 brane setup as in Section \ref{fourD7oneinst_sec} but a different B-field. 
A priori there are rank one instantons contributing to the superpotential originating from $D_1$ and $D_2$. However, if we choose the B-field along $D_2$ to be zero and one-half along $D_1$ this prevents rank-one instantons on $D_2$ while still allowing rank-one instantons on $D_1$ \cite{Berglund:2012gr}. Hence, the leading contribution to the non-perturbative superpotential from $D_2$ are rank-two instantons and
\begin{equation}
 \beta = c \frac{a_2}{a_1} = 2 c\,.
\end{equation}
Thus, we have to look for values of $c \lesssim 1/2$ in order to realize $\beta \lesssim 1$ in our scan. On the other hand, when we choose $B = \frac12 D_2$ we forbid rank-one instantons on $D_1$ such that rank-two instantons on $D_1$ and rank-one instantons on $D_2$ are the leading contributions to the superpotential. In this case, we are looking for $c\lesssim 2$ since $\beta = c/2$.

\begin{figure}[t!]
\centering
\includegraphics[width= 0.47\linewidth]{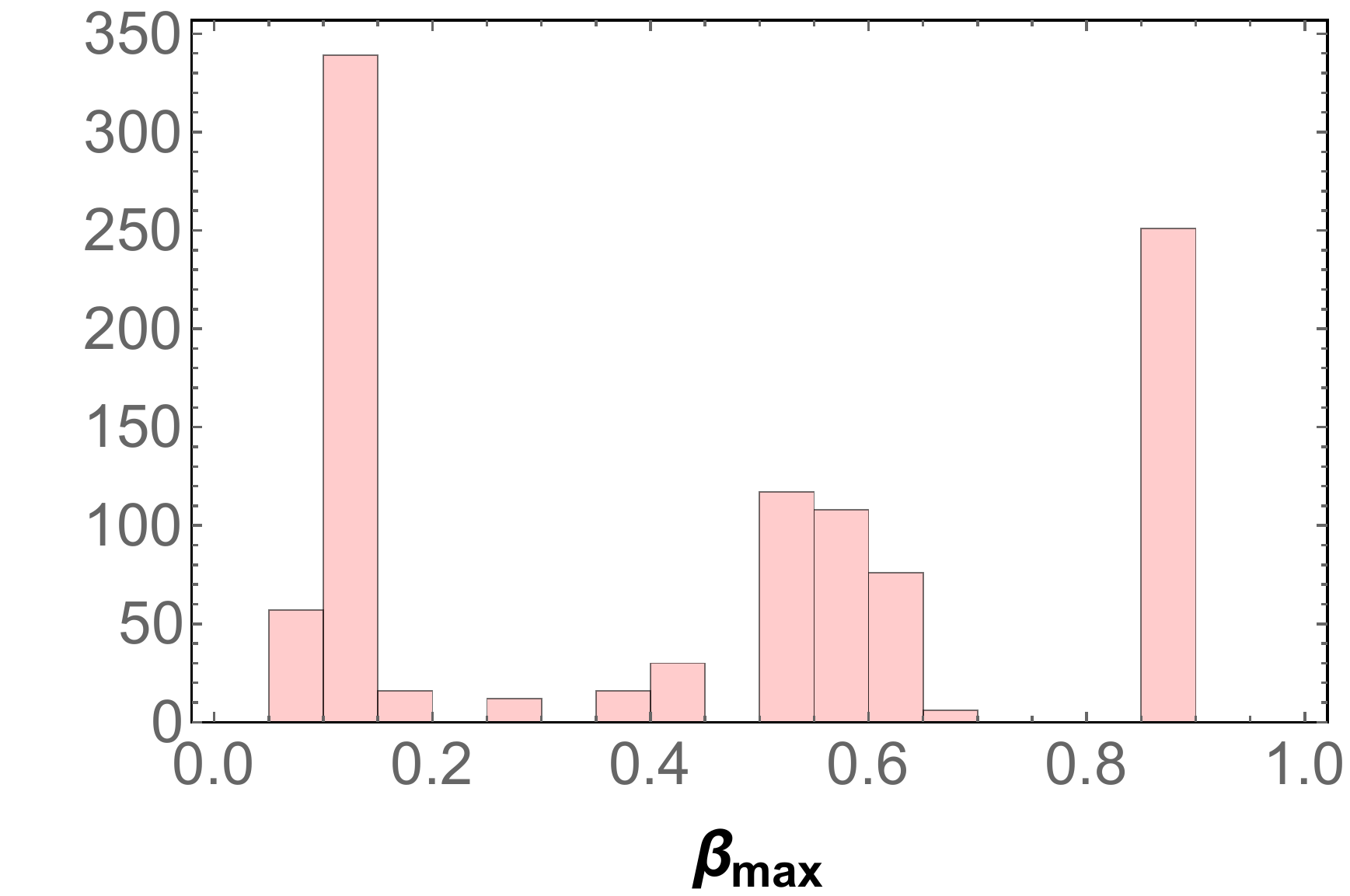}
\includegraphics[width= 0.48\linewidth]{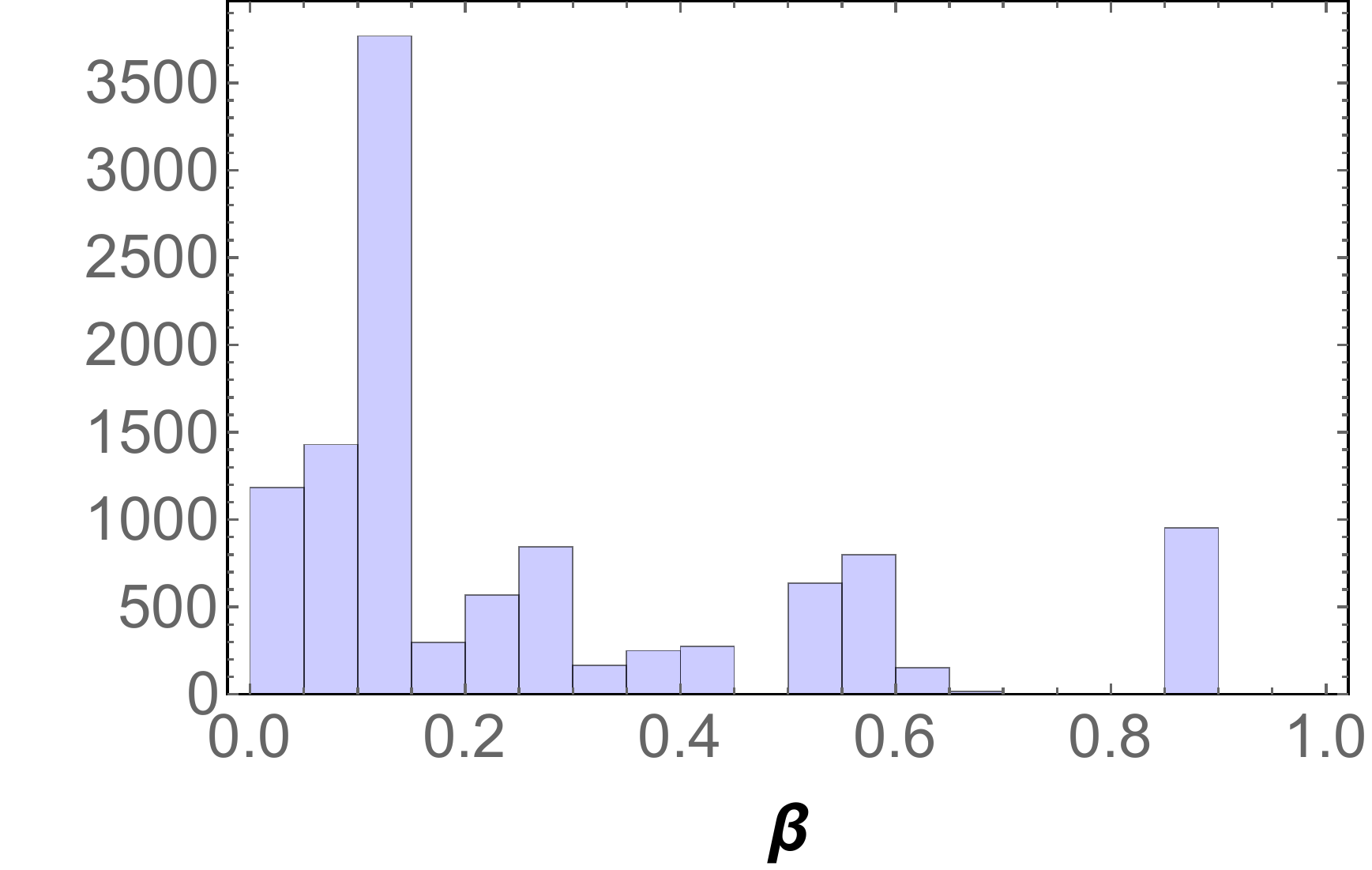}
\caption{Four D7 branes and rank-two instanton: on the left, we show the distribution of $\beta_{\text{max}}$, i.e. the maximal value of $\beta$ for each polytope that we find in our scan. On the right we show all possible values of $\beta$ and their relative distribution in our scan.}
\label{betadistr_fig3}
\end{figure}

The results are presented in Table \ref{sec4res3_tab} and Figure \ref{betadistr_fig3}. The proportion of models that work are somewhat better than those in section \ref{fourD7oneinst_sec} but not as good as in \ref{oneD7oneinst_sec}. This is understandable since the most restricting constraint is, as in section \ref{fourD7oneinst_sec}, the vanishing of $\mathcal{F}_D$ along any other components than $D_1$ and $D_2$. The closest values to one of $\beta$ in this scenario are $8/9, 25/32, 2/3,32/49, 16/25$.

\begin{table}[ht!]
\centering
  \begin{tabular}{|c|c|c|c|}
  \hline
  $h^{1,1}$ & $\# \text{polytopes}$ & $\# \text{polytopes where geometric}$ & $\# \text{polytopes where geometric conditions are}$\\
   & & $\text{conditions are met}$ & $\text{met and suitable fluxes can be found}$\\
  \hline
  $3$ & 244 &  39 &  31\\
  \hline
  $4$ & 1197 &  285 & 246 \\
  \hline
  \end{tabular}
  \caption{Four D7 branes and rank-two instanton: number of polytopes where the D-term generated racetrack uplift can be applied.}
  \label{sec4res3_tab}
\end{table}

\subsection{One D7 brane and rank-two instantons} \label{oneD7twoinst_sec}

\begin{figure}[t!]
\centering
\includegraphics[width= 0.47\linewidth]{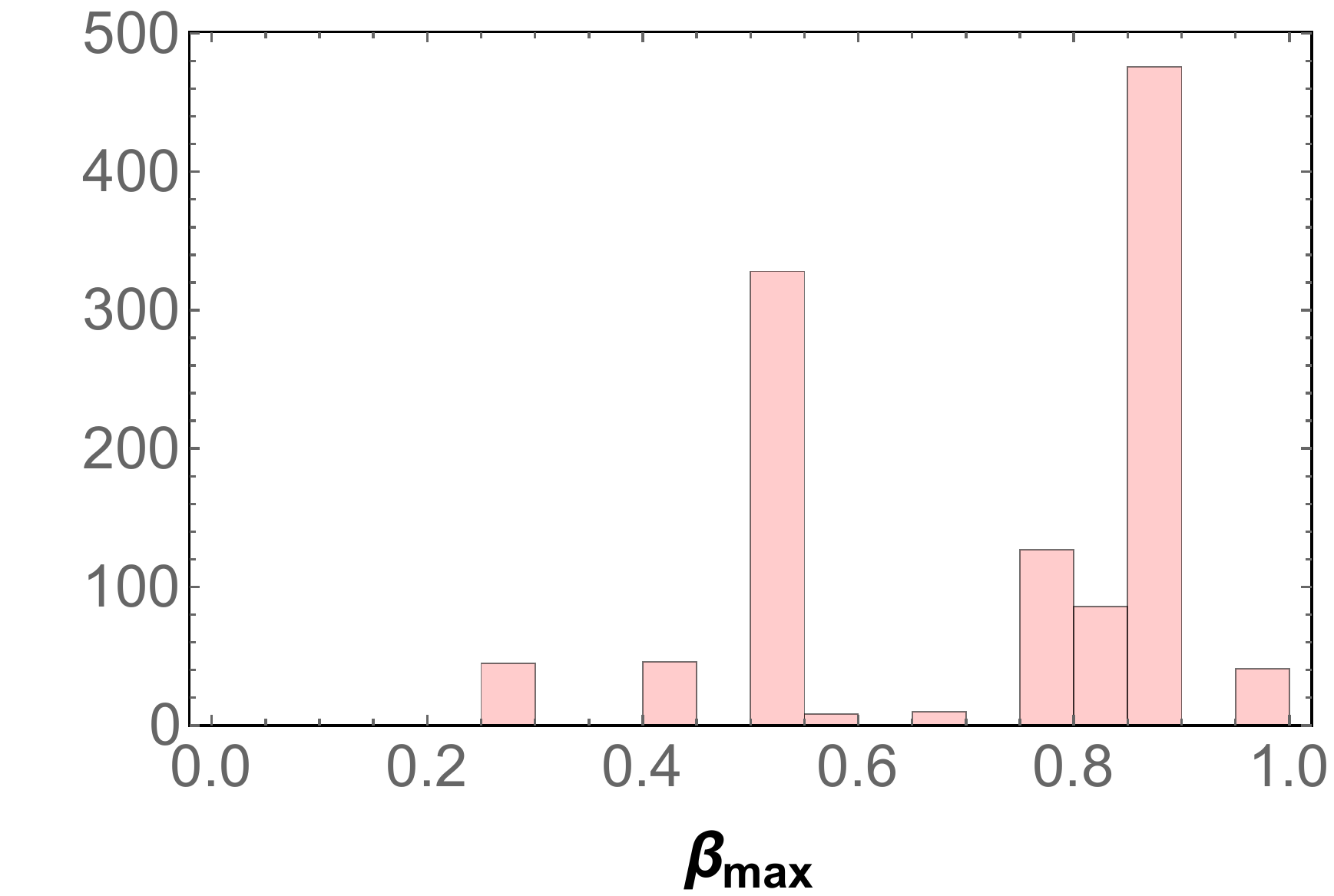}
\includegraphics[width= 0.48\linewidth]{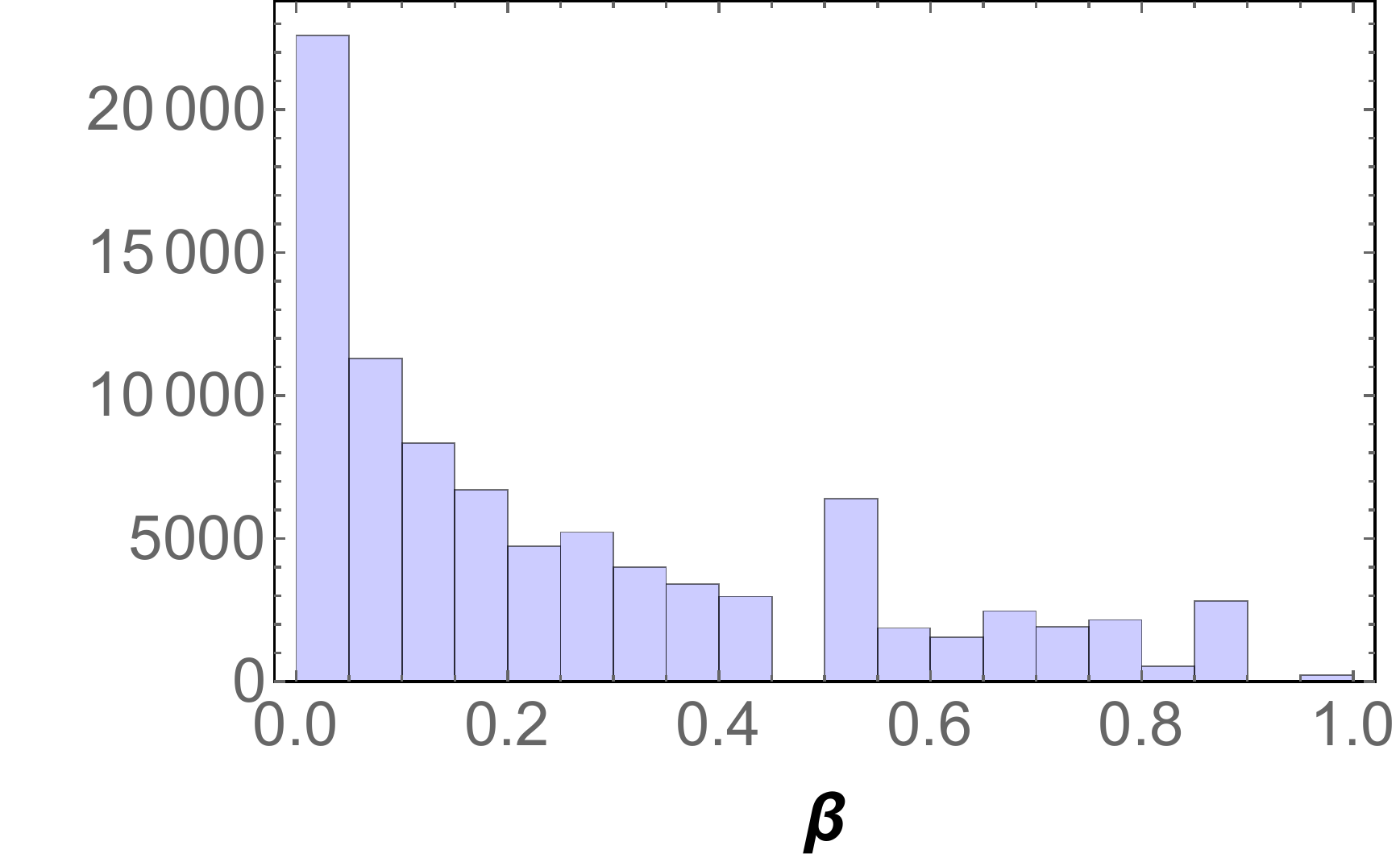}
\caption{One D7 brane and rank-two instanton: on the left, we show the distribution of $\beta_{\text{max}}$, i.e. the maximal value of $\beta$ for each polytope that we find in our scan. On the right we show all possible values of $\beta$ and their relative distribution in our scan.}
\label{betadistr_fig4}
\end{figure}

Finally, we can combine the scenarios of \ref{oneD7oneinst_sec} and \ref{fourD7twoinst_sec}. We consider one D7 brane wrapping the divisor $4 D_D$ and a B-field that forbids rank-one instantons on $D_1$ or $D_2$. The results are presented in Table \ref{sec4res4_tab} and Figure \ref{betadistr_fig4}. We find a similar abundance of working $\beta$ values as in Section \ref{oneD7oneinst_sec}. The values closest to one for $\beta$ are $49/50, 121/128, 8/9, 225/256, 196/225$. This scenario combines the advantages of a possible flux choice for most of the polytopes that allow the geometric conditions of the D-term generated racetrack observed in Section \ref{fourD7twoinst_sec}, and a large variety in $\beta$ values, Section \ref{oneD7oneinst_sec}.

\begin{table}[ht!]
\centering
  \begin{tabular}{|c|c|c|c|}
  \hline
  $h^{1,1}$ & $\# \text{polytopes}$ & $\# \text{polytopes where geometric}$ & $\# \text{polytopes where geometric conditions are}$\\
   & & $\text{conditions are met}$ & $\text{met and suitable fluxes can be found}$\\
  \hline
  $3$ & 244 & 39 &  36\\
  \hline
  $4$ & 1197 & 285 & 230\\
  \hline
  \end{tabular}
  \caption{One D7 branes and rank-two instanton: number of polytopes where the D-term generated racetrack uplift can be applied.}
  \label{sec4res4_tab}
\end{table}

\section{Concrete CY compactifications with D-term generated racetrack uplift}\label{Sec:Examples}

In this section we choose in the list of the models selected by the scan and work out the details of the geometry, the D-brane configurations and the scalar potential, showing how the mechanism works in concrete examples. We start from the simplest example, i.e. two rank one E3-branes wrapping $D_1$ and $D_2$ and an $SO(8)$ stack wrapping the fixed point locus (i.e. $D_D=[O7]$). Scanning among such simple models, we do not find values of $\beta$ close to one. The biggest at $h^{1,1}=3$ we find is $\beta=1/2$. This of course will require some tuning on the prefactor $A_2$ in the non-perturbative superpotential, to make the two contribution roughly of the same order. In the second and in the third models that we present, we show how to construct a slightly more refined model that allows $\beta$ close to one.

\subsection{Example 1: rank one E3-instantons and $SO(8)$ stack}

\subsubsection*{Geometric data}

We choose `geometry ID \# 257' from \cite{Altman:2014bfa} as an example. This can be represented as a CY hypersurface in the toric ambient space described by the following weight system
\begin{align}%
\centering
\begin{tabular}{|c|c|c|c|c|c|c||c|}
\hline
$z_0$  & $z_1$  & $z_2$  & $z_3$  & $z_4$  & $z_5$  & $z_6$  & $eq_{X_3}$ \\
\hline \hline
$1$  & $1$  & $0$  & $0$  & $1$  & $1$  & $4$  & $8$  \\ \hline
$0$  & $0$  & $1$  & $1$  & $0$  & $1$  & $3$  & $6$  \\ \hline
$0$  & $0$  & $1$  & $0$  & $1$  & $0$  & $2$  & $4$  \\
\hline
\end{tabular}
\label{table weight matrix}
\end{align}
and the Stanley-Reisner ideal\footnote{This Stanley-Reisner ideal originates from a specific triangulation of the relevant polytope. The other triangulations gives rise to other patches of the K\"ahler cone of the CY hypersurface.}
\begin{align}
SR= \{z_2z_3, \, z_3z_5,\,z_2z_4z_6, \, z_0z_1z_5, \, z_0z_1z_4z_6 \}\,\,.
\end{align}
The Calabi-Yau three-fold $X_3$ is determined by the zero locus of the polynomial $eq_{X_3}$, whose degrees are specified in the last column in \eqref{table weight matrix}. This CY has $h^{1,1}=3$ and $h^{1,2}=103$ (so that $\chi(X_3)=-200$).

The two rigid divisors with $h^{1,0}=0$ $D_1$ and $D_2$ that will be wrapped by the E3-branes are the toric divisors $D_{z_3}$ and $D_{z_2}$ respectively. The third divisor completing them to a diagonal basis is $D_b = 2 D_{z_3} + D_{z_4}$. The intersection form is
\begin{equation}\label{I3Ex1}
 I_3 = D_1^3 + 2 D_2^3 + 2 D_b^3 \:.
\end{equation}

The K\"ahler form $J$ can be expanded in the diagonal basis as $J = t_b D_b + t_1 D_1 + t_2 D_2$. The volume form is then given by 
\begin{align}
 \begin{aligned}
 \mathcal{V} &= \frac16 \int_{X_3}J\wedge J\wedge J= \frac16 \left(2 t_b^3 +  t_1^3 + 2  t_2^3 \right)\,=\, \frac{1}{3} \tau_b^{3/2} - \frac{\sqrt{2}}{3}\tau_1^{3/2} -  \frac{1}{3} \tau_2^{3/2}\:,
 \end{aligned}
\end{align}
where $\tau_i=\frac12 \int_{D_i} J\wedge J$. The K\"ahler cone condition for the CY three-fold is given as\footnote{This has been obtained by joining the K\"ahler cones of different triangulations of the ambient space that lead to the same Calabi-Yau topology.}
\begin{equation}
 2t_b +t_1 + 2t_2> 0\,,\qquad t_1 <0\,,\qquad t_2 <0\:.
\end{equation}

The toric divisors that intersect both $D_1$ and $D_2$ are $D_{z_0}$, $D_{z_1}$ and $D_{z_6}$; the first two have $\chi(D_{z_0})=\chi(D_{z_1})=25$, while the third has $\chi(D_{z_6})=215$. We choose the orientifold involution to be
\begin{equation}\label{OrInvol}
 z_6 \mapsto - z_6 \:.
\end{equation}
The orientifold invariant equation defining the Calabi-Yau $X_3$ is then:
\begin{equation}
 z_6^2 = h_{8,6,4}(z_0,...,z_5) \:.
\end{equation}

\subsubsection*{D-brane setup}

The O7-plane is located at the codimension one fixed point locus of \eqref{OrInvol}, i.e. at $z_6=0$, and has large negative D3-charge. In terms of the diagonal basis 
\begin{equation}
D_D \equiv D_{z_6} = 4D_b-3D_1-2D_2 \:.
\end{equation}
There are also fixed points: looking at the scaling relations, one can work out that they are located at $z_0=z_1=z_3=0$, $z_2=z_3=z_5=0$ and $z_3=z_4=z_5=0$. The first locus counts one point in the CY $X_3$, while the other two are empty (this is obtained by expressing the loci as the intersection of the three respective toric divisors and computing such intersection by using \eqref{I3Ex1}). Hence we have found one O3-plane.

To cancel the D7-tadpole introduced by the O7-plane, we put four D7-branes plus their four orientifold images on top of the O7-plane on the divisor $D_D=D_{z_6}$. Moreover, there will be E3-instantons wrapping the rigid divisors $D_1=D_{z_3}$ and $D_2=D_{z_2}$. The choice of B-field that allows to have rank one invariant instantons and D7-brane flux with no components along $D_b$ is
\begin{equation}
 B = \frac{D_1}{2} + \frac{D_2}{2} \:.
\end{equation}
This allows zero flux on the E3-instantons, $\cF_1=\cF_2=0$ and the following flux on the D7-brane\footnote{$D_D=D_{z_6}$ is equal to $D_1=D_{z_3}$ mod an even four-cycle. Hence $F_D$ is integral up to $\frac{D_1}{2}$ and the flux $\cF_D=F_D-B$ is then integral up to $\frac{D_2}{2}$.}
\begin{equation}\label{FluxEx1}
 \cF_D = D_1 - \frac32 D_2 \:.
\end{equation}
This flux breaks the $SO(8)$ gauge group to $U(4)$ (actually the diagonal $U(1)$ gets a mass by the St\"uckelberg mechanism, but remains at low energy as a global symmetry).
The D3-charge given by the D7-branes, the O7-plane and the O3-plane is
\begin{equation}
 Q_{D3}= -\frac{N_{O3}}{2} - \frac{\chi(D_D)}{2} - 4 \int_{D_D}\cF_D\wedge \cF_D = - \frac12 - \frac{215}{2} + 48 = -60 \:.
\end{equation}

We now compute the open string spectrum in this setup. First of all we compute how many instanton zero modes we have. They will be in the fundamental representation of the unbroken $U(4)$ gauge group on the D7-brane stack. We also keep track of the $\mathbb{Z}_2$ charge on the invariant E3-brane. The actual calculations are reported in Appendix \ref{sect:zeromodesdetails} and uses the formulae \eqref{ExtE3D7}. Here we present the results. The number of fermion zero modes of the instanton wrapping $D_1$, localized on the curve $\mathcal{C}_{E1}=D_1\cap D_D$ are
\begin{eqnarray}
 N_{{\bf \bar{4}}_{-1},+}^{E3_1} &=& \dim H^1(\mathcal{C}_{E1}, [\cF_D]\otimes [D_1]^{1/2}\otimes [D_D]^{1/2}) = 4 \\
 N_{{\bf 4}_{+1},-}^{E3_1} &=& \dim H^0(\mathcal{C}_{E1}, [\cF_D]\otimes [D_1]^{1/2}\otimes [D_D]^{1/2}) = 1 \:,
\end{eqnarray}
with a chiral index $I_{D7E3}=\int_{D_{z_6} \cap D_1}\cF_D =-3$. We use the symbol $[\cF_D]$ for the line bundle with first Chern class equal to $\cF_D$.\footnote{It may appear strange to use of the gauge invariant combination $\cF_D=F_D-i^\ast B$, as the $B$ field should not contribute to the matter multiplicity. In fact, here $\cF_D=(F_D-i^\ast B)-(F_{E3}-i^\ast B)=F_D-F_{E3}$.}
As explained in Appendix \ref{sect:zeromodesdetails}, $H^0(\mathcal{C},\cL)$ is computed by counting holomorphic sections of the line bundle $\cL$ on $\mathcal{C}$ (i.e. in this case polynomials of given degree). By Serre duality, $H^1(\mathcal{C},\cL)=H^0(\mathcal{C},\cL^{-1}N_{\mathcal{C}})$, and we compute it again by counting holomorphic sections (of a different line bundle).

The same kind of computations lead to the following numbers for the charged zero modes of the E3-instanton wrapping $D_2$:
\begin{eqnarray}
 N_{{\bf \bar{4}}_{-1},+}^{E3_2} &=& \dim H^1(\mathcal{C}_{E2}, [\cF_D]\otimes [D_2]^{1/2}\otimes [D_D]^{1/2}) = 0 \\
 N_{{\bf 4}_{+1},-}^{E3_2} &=& \dim H^0(\mathcal{C}_{E2}, [\cF_D]\otimes [D_2]^{1/2}\otimes [D_D]^{1/2}) = 6 \:,
\end{eqnarray}
where now $\mathcal{C}_{E2}=D_2\cap D_D$. The chiral index is $I_{D7E3}=\int_{D_{z_6} \cap D_1}\cF_D =6$.

The gauge flux $\cF_D$ generates also a chiral spectrum on the worldvolume of the D7-branes. The states are in the antisymmetric representation of $U(4)$ and in its conjugate representation. Their numbers are
\begin{eqnarray}
 N_{{\bf 6}_{+2}}^{D} &=& \dim H^1(D_D, [\cF_D]^{2})+\dim H^0(D_D,[\cF]^2\otimes [D_D])  \geq 9 \\
 N_{{\bf \bar{6}}_{-2}}^{D} &=& \dim H^0(D_D,[\cF]^{-2}\otimes [D_D]) + \dim H^1(D_D,[\cF]^{2}\otimes [D_D]) \geq 15 \:,
\end{eqnarray}
with a chiral index $I_{D7D7'}=2\int_{X_3} D_{z_6}^2\cF_D = - 6$. We have used Serre duality to relate the second cohomology group to the zero cohomology (allowing us to count again holomorphic sections).
In this case, we cannot give a definite number for the dimensions of the extension groups. This happens because we cannot simply count the holomorphic section on $D_D$ by counting the holomorphic sections on the Calabi-Yau that do not vanish identically on $D_D$ (see Appendix \ref{sect:zeromodesdetails}). Anyway, we are able to prove that both numbers are different from zero, which is the result we need out of this computation.

From this spectrum we see that the non-perturbative superpotential is not obstructed. In $S_{\rm inst}$ there will be terms like $\eta_{{\bf 4}_{+1},-} \cdot \phi_{{\bf \bar{6}}_{-2},0}\cdot \eta_{{\bf 4}_{+1},-} $ and $\eta_{{\bf \bar{4}}_{-1},+} \cdot \phi_{{\bf 6}_{+2},0} \cdot \eta_{{\bf \bar{4}}_{-1},+} $
that are gauge invariant and are proper bilinears of the zero modes to get a non-zero contribution after integration on the Grassmann variables (remember that ${\bf 6}$ is the antisymmetric representation and that the second sign is relative to a $\mathbb{Z}_2$ charge).

\subsubsection*{De Sitter minimum}

The flux \eqref{FluxEx1} generates the following FI-term:
\begin{equation}
\zeta_D \,=\, \frac{1}{\mathcal{V}}\int_{D_D}J\wedge \cF_D \,\,\propto\,\, t_1 - 2 t_2\:. 
\end{equation}
Requiring $\zeta_D=0$ implies
\begin{equation}
 \tau_2 \,=\, t_2^2 \,=\, \frac14 t_1^2 \,=\, \frac12 \tau_1 \qquad \Rightarrow \qquad c=\frac12\:.
\end{equation}
Since $a_1=a_2$ in this example, we have 
\begin{equation}
\beta= c \frac{a_2}{a_1} = \frac12 \:.
\end{equation}

We are now ready to calculate the LVS AdS minimum and D-term generated racetrack uplift minimum. We choose (i.e. we assume that after stabilising the complex structure moduli and the dilaton the effective action parameters are fixed at the following values for some choice of flux quanta)
\begin{equation}
 W_0 = 1\,, \qquad \hat \xi = 2.1\,, \qquad A_1 = -0.1 \:.
\end{equation}
If $A_2$ is zero (this can happen if some matter field appearing in the prefactor is fixed to zero VEV), one finds a classical AdS LVS minimum with 
\begin{equation}
 \langle \mathcal{V} \rangle = 4.5 \cdot 10^4\,, \qquad \langle \tau_s \rangle =1.76\,, \qquad \langle V \rangle = - 7.9 \cdot 10^{-16}\,.
\end{equation}

If $A_2$ is non zero, the D-term generated racetrack mechanism takes place. Since we have a value for $\beta$ that is not close to one, we would need to use the flux parameters to have a small values for $A_2$. We take $A_2  =  10^{-4}$; then the D-term generated racetrack de Sitter vacuum is found at~\footnote{For the parameter sets $A_2 = 10^{-3}$ and $\hat \xi = 0.77$ a meta stable de Sitter minimum would be achieved at $\langle \mathcal{V} \rangle = 1.2 \cdot 10^3$ and for $A_2 = 10^{-5}$ and $\hat \xi = 2.8$ at $\langle \mathcal{V} \rangle = 1.0 \cdot 10^6$.}
\begin{equation}
 \langle \mathcal{V} \rangle = 1.4 \cdot 10^5\,, \qquad \langle \tau_s \rangle = 2.00\,, \qquad \langle V \rangle = 2.7 \cdot 10^{-16}\,.
\end{equation}
Note that even though $A_2$ is chosen rather small, it is generally the leading instanton contribution to the superpotential. Higher rank instantons wrapping the small cycles would have prefactors (in absence of tuning for them as well) $A_{1n} e^{-n a_1 \tau_s} \sim A_{1n} e^{-12 n}$ that are quite suppressed for $n \geq 2$ (the suppression factor is bigger than the one coming from the chosen $A_2$). Furthermore, any other instanton effects will have to involve the big cycle $\tau_b$ and are hence completely negligible for the obtained value of the volume.

\subsection{Example 2: rank one E3-instantons and brane/image-brane}

This example is just a simple modification of the previous one. We keep the same CY three-fold \eqref{table weight matrix} and E3-instantons, but change the D7-brane configuration.

\subsubsection*{D-brane setup}

We keep the orientifold involution $z_6 \mapsto - z_6$ and so we have again one O7-plane on the locus $\{z_6=0\}$ and one O3-plane at $z_0=z_1=z_3=0$.
To saturate the D7-brane tadpole we chose a set of one brane wrapping a divisor in the class $4D_{z_6}$ and its image:
\begin{equation}
 D7 \,\, : \qquad \eta+z_6 \psi \qquad \mbox{and} \qquad   D7' \,\, : \qquad \eta-z_6 \psi
\end{equation}
with e.g.
\begin{eqnarray}
 \eta &\equiv &  P_{12}(z_0,z_1)z_2^8z_5^4 + P_{8}(z_0,z_1) z_3^{12}z_4^8 + P_{4}(z_0,z_1)z_2^4z_4^4z_5^8 \:,  \\
 \psi &\equiv &  P_{9}(z_0,z_1)z_2^6z_5^3 + P_{6}(z_0,z_1) z_3^{9}z_4^6 + P_{3}(z_0,z_1)z_2^3z_4^3z_5^6   \:,
\end{eqnarray}
that realize a connected locus for the D7-brane. $P_k(z_0,z_1)$ are homogeneous polynomials of degree $k$ in $z_0,z_1$. We see that $[\eta]$ is in the class $4D_{z_6}$ and $[\psi]$ in $3D_{z_6}$.

The divisor wrapped by the D7-brane is even and the gauge flux $F_D$ is an integral two form. The B-field remains $B=\frac{D_1}{2}+\frac{D_2}{2}$ in order to allow rank one E3-instantons on the two rigid divisors. Hence, we can take the gauge invariant flux on the D7-brane as
\begin{equation}\label{FluxEx2}
 \cF_D = \frac{D_1}{2}-\frac{D_2}{2} \:.
\end{equation}

The D3-charge induced by the D7-branes now changes, according to \eqref{O71D7QD3}:
\begin{equation}
 Q_{D3}= -\frac{N_{O3}}{2} - \frac{\chi(D_{z_6})}{6} - \frac{\chi(4D_{z_6})}{12} - 4 \int_{D_{z_6}}\cF_D\wedge \cF_D = - \frac12 - \frac{215}{6} - \frac{1490}{3} + 7 = -526 \:.
\end{equation}

\

We now compute the open string spectrum. In this example, all the states of interest live on curves. While in the previous example, we were able to compute the actual number of states for each curve, here we cannot. The reason is that the curves cannot be written as hypersurfaces in a toric two-dimensional space.\footnote{Furthermore, the curves are intersections of a high degree divisor, $D_D$ with another divisor. This divisor also supports fluxes that are not of pull-back type (whose non-trivial Poincar\'e dual two cycles in $D_D$ are trivial in the Calabi-Yau $X_3$) and that can change the number of vector-like pairs.} 
We then compute only the index for the intersection of the D7-brane with the E3-brane and give a positive lower bound for both chiralities of the states at the intersection of the D7-brane with its image.

We start from the instanton zero modes. They will be charged under the $U(1)$ symmetry living on the D7-brane. 
The difference of the numbers of fermion zero modes of the instanton wrapping $D_1$, localized on the curve $\mathcal{C}_{E1}=D_1\cap D_D$ is
\begin{equation}
 N_{+1,-}^{E3_1} -  N_{-1,+}^{E3_1} = I_{D7E3_1}=\int_{D_{D} \cap D_1}\cF_D =-6 \:.
\end{equation}
The same kind of computations leads to the following difference of the numbers for the charged zero modes of the E3-instanton wrapping $D_2$:
\begin{equation}
N_{+1,-}^{E3_2} - N_{-1,+}^{E3_2} = I_{D7E3_2}=\int_{D_{D} \cap D_2}\cF_D =8\,.
\end{equation}

The gauge flux $\cF_D$ generates also a chiral spectrum at the intersection of the D7-brane with its image. This curve is given by $\mathcal{C}=\{\eta=\psi=0\}$. The states have charge $\pm2$ with respect to the D7-brane $U(1)$ group. As explained in Appendix \ref{sect:zeromodesdetails}, we can give lower bound for their numbers:
\begin{eqnarray}
 N_{{-2}}^{D} &=& \dim H^0(\mathcal{C}, [\cF_D]^{-2}\otimes [D_{z_6}]^{3}) \geq 295 \\
 N_{{+2}}^{D} &=& \dim H^0(\mathcal{C}, [\cF_D]^2\otimes [D_{z_6}]^{3}) \geq 301 \:.
\end{eqnarray}
Their chiral index is $I_{D7D7'}=12\int_{X_3} D_{z_6}^2\cF_D = 6$. Hence we have fields of both charges under the $U(1)$ of the D7-brane.

Also in this example, the non-perturbative superpotential is not obstructed. First of all, we have fields $\phi$ with both charges, and fields that are neutral: they are counted by $h^{0,2}(D_{z_6})=24$.
It is also important that the difference of the instanton zero modes is an even number, because then also the sum is even as well. This allows $S_{\rm inst}$ to have terms like $\eta_{{+1},-} \cdot \phi_{{-2},0}\cdot \eta_{{+1},-} $, $\eta_{{-1},+} \cdot \phi_{{+2},0} \cdot \eta_{{-1},+} $ and $\eta_{{-1},+} \cdot \phi_{0,0} \cdot \eta_{{+1},-} $
that are gauge invariant and are proper bilinears of the zero modes to get a non-zero contribution after integration on the Grassmann variables.

\subsubsection*{De Sitter minimum}

The flux \eqref{FluxEx2} generates the following FI-term:
\begin{equation}
\xi_D \,=\, \frac{1}{\mathcal{V}}\int_{D_D}J\wedge \cF_D \,\,\propto\,\, 3 t_1 - 4 t_2\:. 
\end{equation}
Requiring $\xi_D=0$ implies
\begin{equation}
 \tau_1 \,=\, t_1^2 \,=\, \frac{16}{9} t_1^2 \,=\, \frac{8}{9} \tau_1 \qquad \Rightarrow \qquad c'=\frac89\:,
\end{equation}
where we called $c'=1/c$: this makes the role of $D_1$ and $D_2$ exchange for this example.
Since $a_1=a_2$ in this example, we have 
\begin{equation}
\beta= c'\frac{a_1}{a_2} = \frac89 \:.
\end{equation}
We notice that now $\beta$ has a value close to one.

We are now ready to calculate the LVS AdS minimum and D-term generated racetrack uplift minimum. We choose
\begin{equation}
 W_0 = 1\,, \qquad \hat \xi = 1.5\,, \qquad A_1 = -0.1 \:.
\end{equation}
If $A_2$ is zero, the AdS LVS minimum is 
\begin{equation}
 \langle \mathcal{V} \rangle = 3.4 \cdot 10^4 \,, \qquad \langle \tau_s \rangle =1.77 \,, \qquad \langle V \rangle = -1.3 \cdot 10^{-15}  \,.
\end{equation}

If $A_2$ is non zero, the D-term generated racetrack mechanism takes place for an order one value of $A_2$. We take $A_2= 5 \cdot 10^{-3} $ and we obtain~\footnote{For the parameter sets $A_2 = 6 \cdot 10^{-3}$ and $\hat \xi = 0.75$ a meta stable de Sitter minimum would be achieved at $\langle \mathcal{V} \rangle = 1.5 \cdot 10^3$ and for $A_2 = 2.5 \cdot 10^{-3}$ and $\hat \xi = 2.1$ at $\langle \mathcal{V} \rangle = 1.5 \cdot 10^6$.}
\begin{equation}
 \langle \mathcal{V} \rangle = 1.3 \cdot 10^5 \,, \qquad \langle \tau_s \rangle = 2.01 \,, \qquad \langle V \rangle = 8.2 \cdot 10^{-17} \,.
\end{equation}

\subsection{Example 3: rank two E3-instanton}

In this second example, we take a CY that would allow only $c=1/2$ or smaller when taking rank one instantons and we choose the B-field to have only rank two instantons on one of the two rigid divisors. This changes the ratio $\frac{a_1}{a_2}$, making $\beta$ closer to one.

\subsubsection*{Geometric data}

We choose the `geometry ID \# 258' from \cite{Altman:2014bfa}. This CY $X_3$ is a hypersurface in the toric ambient space defined by
\begin{align}%
\centering
\begin{tabular}{|c|c|c|c|c|c|c||c|}
\hline
$z_0$  & $z_1$  & $z_2$  & $z_3$  & $z_4$  & $z_5$  & $z_6$  & $eq_{X_3}$ \\
\hline \hline
$0$  & $1$  & $1$  & $2$  & $0$  & $3$  & $7$  & $14$  \\ \hline
$0$  & $0$  & $1$  & $1$  & $1$  & $1$  & $4$  & $8$  \\ \hline
$1$  & $0$  & $0$  & $0$  & $0$  & $1$  & $2$  & $4$  \\
\hline
\end{tabular}
\label{table weight matrix3}
\end{align}
and the Stanley-Reisner ideal 
\begin{align}
SR= \{z_0z_4, \, z_2z_4, \, z_0z_5z_6, \, z_1z_2z_3, \, z_1z_3z_5z_6 \}\,\,.
\end{align}
The Calabi-Yau three-fold is determined by the zero locus of the polynomial $eq_{X_3}$, whose degrees are in the last column in \eqref{table weight matrix3}. As the CY considered previously, it has $h^{1,1}=3$ and $h^{1,2}=103$ (with $\chi(X_3)=-200$). Computing e.g. the intersection numbers for the generators 
of $H^{2}(X_3,\mathbb{Z})$ one sees that this is a topologically different CY manifold from the one considered before, `geometry ID \# 257' from \cite{Altman:2014bfa}.

The two rigid divisors with $h^{1,0}=0$ $D_1$ and $D_2$ that will be wrapped by the E3-branes are the toric divisors $D_{z_0}$ and $D_{z_4}$ respectively. The third divisor completing them to a diagonal basis is $D_b = 2 D_{z_4} + D_{z_5}$. The intersection form is
\begin{equation}\label{I3Ex3}
 I_3 = 9D_1^3 +  D_2^3 + 9 D_b^3 \:.
\end{equation}

The K\"ahler form $J$ can be expanded in the diagonal basis as $J = t_b D_b + t_1 D_1 + t_2 D_2$. The volume form is then given by 
\begin{equation} 
 \mathcal{V} = \frac16 \int_{X_3}J\wedge J\wedge J= \frac16 \left(9 t_b^3 + 9 t_1^3 + t_2^3 \right)\,=\, \frac{1}{9\sqrt{2}} \left( \tau_b^{3/2} - \tau_1^{3/2} -  3 \tau_2^{3/2}\right)\:,
\end{equation}
where $\tau_i=\frac12 \int_{D_i} J\wedge J$. The K\"ahler cone condition is given by
\begin{equation}
 t_b > 0\,,\qquad t_1 <0\,,\qquad t_2 <0\:.
\end{equation}

The toric divisors that intersect both $D_1$ and $D_2$ are $D_{z_1}$, $D_{z_3}$ and $D_{z_6}$; the first two have $\chi(D_{z_0})=13$ and $\chi(D_{z_1})=37$, while the third has $\chi(D_{z_6})=213$. We choose the orientifold involution to be
\begin{equation}
 z_6 \mapsto - z_6 \:.
\end{equation}
The orientifold invariant equation defining the Calabi-Yau $X_3$ is then:
\begin{equation}\label{OrInvolEx3}
 z_6^2 = h_{14,8,4}(z_0,...,z_5) \:.
\end{equation}

\subsubsection*{D-brane setup}

The O7-plane corresponding to \eqref{OrInvolEx3} is located at $z_6=0$ and has large negative D3-charge. In terms of the diagonal basis 
\begin{equation}
D_D \equiv D_{z_6} = \frac73 D_b-\frac13 D_1-3D_2 \:.
\end{equation}
There are also fixed points: they are located at $z_0=z_1=z_2=0$, $z_1=z_2=z_5=0$ and $z_1=z_3=z_4=0$. Each locus is made up of one point in the CY $X_3$. Hence we have found three O3-planes.

To cancel the D7-tadpole introduced by the O7-plane, we put four D7-branes plus their four images on top of the O7-plane on the divisor $D_D=D_{z_6}$. Moreover, there will be E3-instantons wrapping the rigid divisors $D_1=D_{z_0}$ and $D_2=D_{z_4}$. We choose the B-field in such a way that it allows rank one E3 instantons only on $D_1$, while preventing it on $D_2$ (as explained in Section \ref{E3instantons_sec}) and that allows to cancel the components of $\cF_D$ along $D_b$ by a proper choice of $F_D$ (that satisfies Freed-Witten anomaly cancellation):
\begin{equation}
 B = \frac{D_1}{2} + \frac{D_b}{2} \:.
\end{equation}
Then we can have $\cF_1=0$ on the E3-brane wrapping $D_1$ and we can choose\footnote{The D7-brane divisor $D_D$ is equal to $D^{\rm even}+D_1+D_2+D_b$ 
where $D^{\rm even}=4D_{z_1}$ is an even divisor class.
Hence $F_D=F^{\rm int} + \frac{D_1}{2} + \frac{D_2}{2} + \frac{D_b}{2}$ and with the chosen B-field the gauge invariant flux $\cF_D$ must have an half-integral component along $D_2$.} 
\begin{equation} \label{FluxEx3}
 \cF_D = D_1 - \frac{D_2}{2} \:.
\end{equation}

The second E3-instanton will have flux $\cF_2=\frac{D_2}{2}$ on one brane of the stack and $-\cF_2$ on the image brane.

The D3-charge given by the D7-branes, the O7-plane and the O3-plane is
\begin{equation}
 Q_{D3}= -\frac{N_{O3}}{2} - \frac{\chi(D_D)}{2} - 4 \int_{D_D}\cF_D\wedge \cF_D = - \frac32 - \frac{213}{2} + 15 = -93 \:.
\end{equation}

\

We now compute the open string spectrum in this setup. As in Example 1, the instanton zero modes are in the fundamental representation of the unbroken $U(4)$ gauge group on the D7-brane stack.  The actual calculations are reported in Appendix \ref{sect:zeromodesdetails}. As regard the rank one E3 instanton wrapping $D_1$, the computation is similar to Example 1 and the results are:
\begin{eqnarray}
 N_{{\bf \bar{4}}_{-1},+}^{E3_1} &=& \dim H^1(\mathcal{C}_{E1}, [\cF_D]\otimes [D_1]^{1/2}\otimes [D_D]^{1/2}) = 3 \\
 N_{{\bf 4}_{+1},-}^{E3_1} &=& \dim H^0(\mathcal{C}_{E1}, [\cF_D]\otimes [D_1]^{1/2}\otimes [D_D]^{1/2}) = 0 \:,
\end{eqnarray}
with a chiral index $I_{D7E3}=\int_{D_{z_6} \cap D_1}\cF_D =-3$. In fact, this curve is a sphere.

For the rank two instanton, the situation is a bit different. The flux on the instanton keeps the gauge group $SO(2)=U(1)$. There will be four type of states, relative to charges $(\pm,\pm)$ with respect to the D7-brane and E3-brane groups. We split into pairs of conjugate representations:
\begin{eqnarray}
 N_{{\bf \bar{4}}_{-1},+1}^{E3_2} &=& \dim H^1(\mathcal{C}_{E2}, [\cF_D]\otimes [\cF_2]^{-1}\otimes [D_2]^{1/2}\otimes [D_D]^{1/2}) = 0 \\
 N_{{\bf 4}_{+1},-1}^{E3_2} &=& \dim H^0(\mathcal{C}_{E2}, [\cF_D]\otimes [\cF_2]^{-1}\otimes [D_2]^{1/2}\otimes [D_D]^{1/2}) = 3 \:,
\end{eqnarray}
with chiral index $I_{D7E3}=\int_{D_{z_6} \cap D_1}\cF_D-\cF_2=3$ and 
\begin{eqnarray}
 N_{{\bf \bar{4}}_{-1},-1}^{E3_2} &=& \dim H^1(\mathcal{C}_{E2}, [\cF_D]\otimes [\cF_2]\otimes [D_2]^{1/2}\otimes [D_D]^{1/2}) = 2 \\
 N_{{\bf 4}_{+1},+1}^{E3_2} &=& \dim H^0(\mathcal{C}_{E2}, [\cF_D]\otimes [\cF_2]\otimes [D_2]^{1/2}\otimes [D_D]^{1/2}) = 2 \:,
\end{eqnarray}
with chiral index $I_{D7E3}=\int_{D_{z_6} \cap D_1}\cF_D+\cF_2 =0$.

The gauge flux $\cF_D$ generates a chiral spectrum on the worldvolume of the D7-branes. The states are in the antisymmetric representation of $U(4)$ and in its conjugate representation. Their numbers are
\begin{eqnarray}
 N_{{\bf 6}_{+2}}^{D} &=& \dim H^1(D_D, [\cF_D]^{2})+\dim H^0(D_D,[\cF]^2\otimes [D_D]) \geq 18 \\
 N_{{\bf \bar{6}}_{-2}}^{D} &=& \dim H^0(D_D,[\cF]^{-2}\otimes [D_D]) + \dim H^1(D_D,[\cF]^{2}\otimes [D_D]) \geq 25 \:,
\end{eqnarray}
with a chiral index $I_{D7D7'}=2\int_{X_3} D_{z_6}^2\cF_D = -7$.

As for Example 1, these results mean that the non-perturbative superpotential is not obstructed.

\subsubsection*{De Sitter minimum}

The flux \eqref{FluxEx3} generates the following FI-term:
\begin{equation}
\xi_D \,=\, \frac{1}{\mathcal{V}}\int_{D_D}J\wedge \cF_D \,\,\propto\,\, 2t_1 - t_2\:. 
\end{equation}
Requiring $\xi_D=0$ implies
\begin{equation}
 \tau_2 \,=\, \frac{t_2^2}{2} \,=\, 2 t_1^2 \,=\, \frac49 \tau_1 \qquad \Rightarrow \qquad c=\frac49\:.
\end{equation}
In this case it is no more true that $a_1$ and $a_2$ are the same, but $a_2=2a_1$. Hence
\begin{equation}
\beta= c \frac{a_2}{a_1} = \frac89 \:.
\end{equation}
We have seen another way to get a $\beta$ close to one.

We are now ready to calculate the LVS AdS minimum and D-term generated racetrack uplift minimum. We choose
\begin{equation}
 W_0 = 1 \,, \qquad \hat \xi = 2.1 \,, \qquad A_1 = -0.1 \:.
\end{equation}
If $A_2$ is zero, one finds a classical AdS LVS minimum with 
\begin{equation}
 \langle \mathcal{V} \rangle = 4.5 \cdot 10^4 \,, \qquad \langle \tau_s \rangle = 1.76 \,, \qquad \langle V \rangle = -7.9 \cdot 10^{-16} \,.
\end{equation}

If $A_2$ is non zero, $A_2 = 3\cdot 10^{-3}$, the D-term generated racetrack mechanism takes place with the following results:\footnote{For the parameter sets $A_2 = 6 \cdot 10^{-3}$ and $\hat \xi = 1.1$ a meta stable de Sitter minimum would be achieved at $\langle \mathcal{V} \rangle = 3.8 \cdot 10^3$ and for $A_2 = 2.5 \cdot 10^{-3}$ and $\hat \xi = 2.5$ at $\langle \mathcal{V} \rangle = 1.4 \cdot 10^6$.}
\begin{equation}
 \langle \mathcal{V} \rangle = 1.4 \cdot 10^5 \,, \qquad \langle \tau_s \rangle = 1.96 \,, \qquad \langle V \rangle = 6.9 \cdot 10^{-17} \,.
\end{equation}

\section{Summary and conclusion}\label{Sec:discussion}

In \cite{Rummel:2014raa} a new mechanism for obtaining de Sitter minima in the Type IIB landscape was proposed. In this paper we implemented this proposal in a more detailed setup.  We considered Calabi-Yau hypersurfaces in toric varieties from the Kreuzer-Skarke list \cite{Kreuzer:2000xy} and employed the results
of \cite{Altman:2014bfa} concerning triangulations of polyhedra for small $h^{1,1}$. This allows to be very explicit in the topology of the three-fold and of its divisors. We revisited the necessary conditions for this mechanism and we found a simple setup of branes where they could be realized. First, we need to have a Calabi-Yau with two rigid divisors that do not intersect each other and that can be completed to a basis of divisors that do not intersect them. We implemented a scan in the Kreuzer-Skarke list for $h^{1,1}\leq4$ that produced a reduced list of CYs that satisfy this condition. We chose to work with orientifold involutions with $h^{1,1}_-=0$ that are realized by inverting the sign of one toric coordinate. The scan on toric hypersurfaces also gives candidates for such coordinates, such that the corresponding 
orientifold planes intersect the two rigid divisors and have a large Euler number. This allows to satisfy the D7-brane tadpole with D7-branes suitable to generate the wanted FI-term, and at the same time to generate a large (negative) D3-charge necessary for the large tunability of fluxes.
This reduced list is a nice result by itself. Finding rigid (orientifold invariant) divisors that appear as `small cycles'  in the volume form of the CY is a typical challenge for moduli stabilization. In particular in the LVS, in absence of a tuned D-term that fixes the two sizes to be of comparable sizes, the two non-perturbative contribution could generate a hierarchy between the corresponding K\"ahler moduli, making one of them much lighter than the other. This situation is particularly interesting for inflation, as the lightest modulus could play the role of the inflaton.

Starting from the reduced list of CYs and the candidate divisors for the D7-branes, we scanned over gauge flux and B-field choices that fulfil the D3-tadpole cancellation condition and that allow Freed-Witten anomaly cancellation. We considered simple situations, where the non-perturbative effects were generated by rank-one and rank-two E3-branes and the D-term was generated by four D7-branes on top of the O7-plane or by a single D7-brane in the class $4[O7]$. For each choice of flux, we computed the proportionality factor $\beta$ between the exponents of the non-perturbative superpotential terms. We found several models where this factor is close to one, allowing the realization of the dS uplift mechanism introduced in \cite{Rummel:2014raa}.

To show how the setup can be constructed in the simple configurations we chose, we worked out the details for a couple of models. We computed the topology of the CY manifold and of its relevant divisors, we made a simple choice of flux and B-field, we computed the D3-charge checking that it can be cancelled and we studied the zero mode structure to be sure that the non-perturbative superpotential is not trivially zero. We finally stabilized the K\"ahler moduli explicitly and found a dS minimum of the potential. 

The setup studied is based on assuming the existence of a (so far unknown) open string moduli potential that fixes the matter field contribution to the D-term potential to zero, while keeping some VEV of them different from zero. This is a strong assumption that may or may not be realized for an explicit open string moduli stabilization. A simple situation would be that the VEVs are all fixed to zero, but this would destroy the non-perturbative superpotential if these VEVs appear as a proportionality factor of the coefficient in front of the exponential, as realized in our examples. This problem (that is due to poor control on open string moduli stabilization) can be solved by making the setup more complicated, e.g. by choosing an orientifold projection with $h^{1,1}\neq 0$ \cite{Grimm:2011dj}. In this case one can turn on an odd flux on the instantonic D3-brane that can compensate the flux on the D7-brane at the intersection, without touching the D-term constraint on the K\"ahler moduli. This would eliminate the matter field VEVs in the prefactor of the non-perturbative superpotential and allow to fix them at zero value. This setup would 
keep the features of the D-brane configurations necessary for the uplift mechanism realization, while making the setup more complicated.\footnote{Another way to solve this problem, with the price of complicating the setup, is to consider E3 instantons wrapping non-rigid divisors with the form $D_{E3_1}=D_1 + e_1 D_2$ and $D_{E3_2}=D_1 + e_2 D_3$ with respect to three small divisors $D_{1,2,3}$ (we then need $h^{1,1}(CY)\geq 4$). The neutral zero modes will be lifted by a suitable (trivial) flux \cite{Bianchi:2011qh}.
The FI-term may be given by $\zeta_D \propto t_1 - t_2/e_1 - t_3/e_2$ with $J = t_1 D_1 + t_2 D_2 + t_3 D_3 + \cdots$.
Now the chiral indexes become $I_{D7E3_1} = I_{D7E3_2} = 0$ while keeping the D-term constraint between $\tau_{1,2,3}$.
In this scenario, the two instantons generate the racetrack potential for uplift if the remaining modulus, say $\tau_3$, is stabilized in the other ways, e.g. by string loop corrections.} 
In this paper we chose to live with the strong assumption and leave the more involved setup for the future, as we were interested to show the abundance of models with the required $\beta\lesssim 1$. In the next step, we will try to implement the mechanism in explicit models with $h^{1,1}_-\neq 0$ and with a visible sector. This will require extending the CY scan to larger $h^{1,1}$.

\section*{Acknowledgements}
We would like to thank Michele Cicoli, Andres Collinucci, Luca Martucci and Timo Weigand for valuable discussions and important comments. The research of APB was supported by the STFC grant ST/L000474/1 and the EPSCR grant EP/J010790/1. MR is supported by the ERC grant `Supersymmetry Breaking in String Theory'. This work was partially supported by a grant from the Simons Foundation and by the Grant-in-Aid for Scientific Research (No. 26247042, 23244057) from the Japan Society for the Promotion of Science. This work was completed while RV and APB were at the Aspen Center for Physics, which is supported by National Science Foundation grant PHY-1066293.

\appendix

\section{Matter and instanton zero modes}
\label{sect:zeromodesdetails}

In this appendix, we report the calculations of the zero modes living on the branes or at their intersections. As explained in Section \ref{Sec:DbraneConfig}, what we will do is computing the number of holomorphic sections of a given line bundle on the surface wrapped by the brane or on the intersection curve. Since this is not possible directly, we will estimate this number by relating it to the number of sections of the corresponding line bundle on some toric ambient space. Due to the structure of the toric spaces, it is easy to make the last computation: one simply need to count monomials of a given degree. Most of the time we will be able to relate the cohomology on the divisor $D\subset X_3$ or on the curve $\cC \subset X_3$ to the cohomology on the CY $X_3$. To compute holomorphic sections of a line bundle $\cL$ on $X_3$ is easy in cases we considered, i.e. when the divisors on $X_3$ descend all from the ambient space $X_4$. First of all the line bundle $\cL$ extends to $X_4$ (in this case, there is a one to one correspondence between line bundles on $X_3$ and $X_4$). Second, the holomorphic section of $\cL$ on the CY are the holomorphic section of $\cL$ on $X_4$ up to sections that vanish identically on $X_3$. In practise, the short exact sequence
\begin{equation}
  0 \rightarrow \mathcal{O}_{X_4}(L-X_3) \rightarrow \mathcal{O}_{X_4}(L) \rightarrow \mathcal{O}_{X_3}(L) \rightarrow 0 \:,
\end{equation}
where $\cL=[L]$, produces a long exact sequence at the level of the cohomology groups
\begin{align}
 0 &  \rightarrow  H^0(X_4,\mathcal{O}_{X_4}(L-X_3)) \rightarrow H^0(X_4,\mathcal{O}_{X_4}(L)) \rightarrow H^0(X_3,\mathcal{O}_{X_3}(L)) \rightarrow  \nonumber \\
  & \rightarrow  H^1(X_4,\mathcal{O}_{X_4}(L-X_3)) \rightarrow ...  \:. \nonumber  
\end{align}
We are simply saying that in the cases we are considering (when the divisors on $X_3$ is always the intersection of the CY equation and a divisor on $X_4$) $H^1(X_4,\mathcal{O}_{X_4}(L-X_3))=0$, i.e. there are no non-holomorphic sections on $X_4$ that restric to holomorphic sections on $X_3$.

\subsection{States localized on surfaces}

This case is relevant for the states localized on the divisor $D_D$, where the four D7-branes and their images live.
As mentioned at the end of Section \ref{Sec:DbraneConfig}, they are counted by ${\rm Ext}^1$ and ${\rm Ext}^2$ in \eqref{StatesMagnBr}. By using the Hirzebruch-Riemann-Roch theorem, Serre duality and the vanishing of $h^0\left(D,E_{a}\otimes E_{b}^\vee \right)$ and $h^2\left(D,E_{a}\otimes E_{b}^\vee \otimes N_{D}\right)$ we can replace the dimension of the first cohomology group by
\begin{eqnarray}\label{h1simplified}
&&h^1\left(D,E_{a}\otimes E_{b}^\vee\right) = h^0\left(D,E_{a}^\vee\otimes E_{b} \otimes N_{D} \right) - \chi\left(D, E_{a}\otimes E_{b}^\vee\right)  \:, \\
&&h^1\left(D,E_{a}\otimes E_{b}^\vee\otimes N_{D}\right) = h^0\left(D,E_{a}\otimes E_{b}^\vee \otimes N_{D} \right) - \chi\left(D, E_{a}\otimes E_{b}^\vee\otimes N_{D}\right)    \:.
\end{eqnarray}
Using these relations, we can express the dimensions of the relevant extension groups as
\begin{eqnarray}\label{dimExtsiplifie}
 \dim\mbox{Ext}^1\left(i_\ast E_{a}, i_\ast E_b\right) &=& h^0\left(D,E_{a}\otimes E_{b}^\vee \otimes N_{D} \right)+h^0\left(D,E_{a}^\vee\otimes E_{b} \otimes N_{D} \right) - \chi\left(D, E_{a}\otimes E_{b}^\vee\right),  \\
 \dim \mbox{Ext}^2\left(i_\ast E_{a}, i_\ast E_b\right) &=& h^0\left(D,E_{a}\otimes E_{b}^\vee \otimes N_{D} \right)+h^0\left(D,E_{a}^\vee\otimes E_{b} \otimes N_{D} \right) - \chi\left(D, E_{a}\otimes E_{b}^\vee\otimes N_{D}\right).\nonumber
\end{eqnarray}
Manifestly their difference gives the index $I_{ab}$. 

One can use these relations to compute the dimensions of the two groups separately. Holomorphic Euler characteristics are easy to compute, thanks to the index theorem $\chi(D,E)=\int_D \mbox{ch}(E)\, \mbox{Td}(D)$. If one is able to count holomorphic sections of the given line bundles, then the formulae \eqref{dimExtsiplifie} give the wanted result.

In case it is not possible to count all of the holomorphic sections, these formulae are still useful to understand what is relevant for us, i.e. knowing if both dimensions are different from zero. In fact, only in this case the non-perturbative superpotential has chances to be generated.
We just need to use the fact that $h^i\geq 0$ and compute the holomorphic Euler characteristic. Remember that in the studied cases $D=D_D$, $E_a=[\cF_D]$ and $E_b=[\cF_D]^{-1}$.

\subsubsection*{Example 3}

In example 3, we have $\chi(D_D,[\cF_D]^2[D_D])=14>0$ and $\chi(D_D,[\cF_D]^2)=21>0$. Hence, equations \eqref{h1simplified} tell us that both $h^0>0$ as well, and one can use this to prove that $\dim$Ext$^i>0$ for both $i=1$ and $i=2$. In fact, these dimensions are the sum of two positive terms and one of these is always one of the two $h^0$ that we have just proven to be positive. Plugging the values for $\chi$ into \eqref{h1simplified}, we estimate
$h^0(D_D,[\cF_D]^{-2}[D_D])\geq 21$ and $h^0(D_D,[\cF_D]^{2}[D_D])\geq 14$. 

If we were able to prove that the long exact sequence is truncated to the short one, as for $X_3\subset X_4$, then we would be able to compute the exact number for the dimensions of the Ext groups, by calculating the number of holomorphic sections of the same line bundle on $X_3$ (mod the sections that identically vanish on $D_D$, that in the present example are absent). 
Luckily, we can show that this computation gives a subset of the wanted holomorphic sections. For our purposes, we only need to prove that these subsets are non-empty. These sections, for $ [\cF_D]^{2}[D_D]$ on $D_D$, are counted by the polynomials of degree $(7,3,4)$ (i.e. sections of $[\cF_D]^{2}[D_D]$ on $X_4$), whose number is $18$. Hence, we actually have $h^0(D_D,[\cF_D]^{2}[D_D])\geq 18$, and since the index is $7$, we improve also the other bound: $h^0(D_D,[\cF_D]^{-2}[D_D])\geq 25$.

\subsubsection*{Example 1}

In the Example 1, this does not happen, unfortunately.  Only $\chi(D_D,[\cF_D]^2[D_D])=4>0$, that implies $h^0(D_D,[\cF_D]^{-2}[D_D])>4$, while $\chi(D_D,[\cF_D]^2)=-2<0$. Here we can use the subset of the holomorphic sections of the line bundle $[\cF_D]^{-2}[D_D]$ that are counted by the polynomials of degree $(4,4,5)$, whose number is $13$. Therefore $h^1([\cF]^2)=-\chi([\cF]^2)+h^0(D_D,[\cF_D]^{-2}[D_D])\geq 9>0$. Hence, we conclude again that $\dim$Ext$^i>0$ for both $i=1$ and $i=2$. In particular we have $\dim$Ext$^1\geq 9$ and $\dim$Ext$^1\geq 15$.

\subsubsection*{Bound from sections on the CY}

The previous considerations were based on the assumption that counting the elements of $H^0(X_3,\mathcal{O}(L))$ gave a lower bound for number of the elements of $H^0(D_D,\mathcal{O}(L))$. We now prove this.

Consider the short exact sequence
\begin{equation}
  0 \rightarrow \mathcal{O}_{X_3}(L-D_D) \rightarrow \mathcal{O}_{X_3}(L) \rightarrow \mathcal{O}_{D_D}(L) \rightarrow 0 \:,
\end{equation}
where $L=\pm 2\cF_D+D_D$.
Its associated long exact sequence of cohomology groups is
\begin{align}
 0 &  \rightarrow  H^0(X_3,\mathcal{O}_{X_3}(L-D_D)) \rightarrow H^0(X_3,\mathcal{O}_{X_3}(L)) \rightarrow H^0(D_D,\mathcal{O}_{D_D}(L)) \rightarrow  \nonumber \\
  & \rightarrow  H^1(X_3,\mathcal{O}_{X_3}(L-D_D)) \rightarrow ...  \:. \nonumber  
\end{align}
The point is that in all our cases, $H^0(X_3,\mathcal{O}_{X_3}(L-D_D))=0$. Hence, the map 
\begin{equation}
H^0(X_3,\mathcal{O}_{X_3}(L)) \rightarrow H^0(D_D,\mathcal{O}_{D_D}(L))\,,
\end{equation}
is injective, implying that $h^0(X_3,\mathcal{O}_{X_3}(L)) \leq h^0(D_D,\mathcal{O}_{D_D}(L))$.

\subsection{States localized on curves}

This is the case for charged instanton zero modes living at the intersection $E3\cap D7$ and matter fields localized at the intersection of the brane and its image $D7\cap D7'$.

The number of states of both chiralities are counted by the first and the second cohomology groups, see \eqref{ExtE3D7} for the intersection $E3\cap D7$.
By Serre duality, the dimension of $H^1(\mathcal{C},\cL\otimes K_{\mathcal{C}}^{1/2})$ is equal to $h^0(\mathcal{C},\cL^{-1}\otimes K_{\mathcal{C}}^{1/2})$ (where $\cL$ is a line bundle). Hence we just need to count holomorphic sections of some line bundle on the curve. 

If we can write the curve as a hypersurface in a two-dimensional toric space, we will show that we can get exact results for the number of zero modes of both chiralities. This happens for the intersections between the E3-branes and the D7-brane stack in Examples 1 and 3. In the Example 2, this is not possible. We will then compute a subset of the holomorphic sections of the given line bundle, which gives a lower bound for the vector like pairs.

\subsubsection*{Example 1 - $\mathcal{C}_{D7\cap E2}$}

We start with the curves at the intersection of the D7-brane stack on $D_D=D_{z_6}$ and the E3-brane wrapping the divisor $D_2=D_{z_2}$. This matter curve is given by setting $z_6=z_2=0$ intersected with the equation defining the CY three-fold. Looking at the SR-ideal, we see that we can set $z_3=1$ and $z_4=1$. This fixes two of the scaling relations. We are left with describing the curve by an equation in a two-dimensional toric space $X_2$:
\begin{align}%
\centering
\begin{tabular}{|c|c|c||c|}
\hline
$z_0$  & $z_1$   & $z_5$  &  $eq_{\mathcal{C}}$ \\
\hline \hline
$1$  & $1$   & $1$   & $4$  \\ 
\hline
\end{tabular}
\end{align}
This is a genus $g=3$ curve, defined by a homogeneous equation of degree $4$ in $\mathbb{CP}^2$. 

We want to count the holomorphic sections of the line bundles $\mathcal{O}(L_i)_{\mathcal{C}}$, where $L_1=\frac{D_D}{2}+\frac{D_2}{2}+\cF_D$ and $L_2=\frac{D_D}{2}+\frac{D_2}{2}-\cF_D$. 
To do this, we start from the exact sequence (it is the structure sequence of $\cC$ twisted by the line bundle $\mathcal{O}(L_i)$)
\begin{equation}
  0 \rightarrow \mathcal{O}_{X_2}(L_i-\mathcal{C}) \rightarrow \mathcal{O}_{X_2}(L_i) \rightarrow \mathcal{O}_{\mathcal{C}}(L_i) \rightarrow 0 \:,
\end{equation}
where $\mathcal{O}(L_i)_{X_2}$ are line bundles defined on the ambient space, i.e. on $\mathbb{CP}^2$. 
From this short exact sequence, we can construct a long exact sequence of cohomology groups:
\begin{align}
 0 &  \rightarrow  H^0(X_2,\mathcal{O}_{X_2}(L_i-\mathcal{C})) \rightarrow H^0(X_2,\mathcal{O}_{X_2}(L_i)) \rightarrow H^0(\mathcal{C},\mathcal{O}_{\mathcal{C}}(L_i)) \rightarrow  \nonumber \\
  & \rightarrow  H^1(X_2,\mathcal{O}_{X_2}(L_i-\mathcal{C})) \rightarrow H^1(X_2,\mathcal{O}_{X_2}(L_i)) \rightarrow H^1(\mathcal{C},\mathcal{O}_{\mathcal{C}} (L_i)) \rightarrow  \nonumber \\
  & \rightarrow  H^2(X_2,\mathcal{O}_{X_2}(L_i-\mathcal{C})) \rightarrow H^2(X_2,\mathcal{O}_{X_2}(L_i)) \rightarrow 0 \:. \nonumber  
\end{align}
If we show that $H^1(X_2,\mathcal{O}_{X_2}(L_i-\mathcal{C}))=0$, then the exact sequence stops there and we have 
\begin{equation}
   h^0 (\mathcal{C},\mathcal{O}_{\mathcal{C}}(L_i)) = h^0(X_2,\mathcal{O}_{X_2}(L_i))  - h^0(X_2,\mathcal{O}_{X_2}(L_i-\mathcal{C})) \:.
\end{equation} 
In our case this actually happens, as we show below. Plugging in the values for the explicit examples, i.e. $L_1=2H$,  $L_2=-H$ and $\cC=4H$, we have
\begin{eqnarray}
     h^0 (\mathcal{C},[D_D]^{1/2}[D_2]^{1/2} [\cF]) = h^0 (\mathcal{C},\mathcal{O}_{\mathcal{C}}(2H)) = 6 - 0 = 6 \:, \\
     h^0 (\mathcal{C},[D_D]^{1/2}[D_2]^{1/2} [\cF]^{-1}) = h^0 (\mathcal{C},\mathcal{O}_{\mathcal{C}}(-H)) = 0 - 0 = 0 \:, 
\end{eqnarray}
where we have used the fact that the number of holomorphic sections of $\mathcal{O}(nH)$ on $\mathbb{CP}^2$ is counted by the homogeneous polynomials of degree $n$. In particular, when $n=2$ we have $6$ polynomials, while when $n$ is negative we have none.

We finish by proving that $H^1(X_2,\mathcal{O}_{X_2}(L_i-\mathcal{C}))=0$. To do this, we again use the Hirzebruch-Riemann-Roch theorem for a two-fold, i.e. $h^0(\cL)-h^1(\cL)+h^2(\cL)=\int_{X_2} \mbox{ch}(\cL) \mbox{Td}(X_2)$. In our case $h^0(X_2,\mathcal{O}_{X_2}(L_1-\mathcal{C})) = h^0(X_2,\mathcal{O}(-2H))=0$ and $h^0(X_2,\mathcal{O}_{X_2}(L_2-\mathcal{C})) = h^0(X_2,\mathcal{O}(-5H))=0$. By Serre duality $h^2(X_2,\mathcal{O}_{X_2}(L_1-\mathcal{C})) = h^0(X_2,\mathcal{O}_{X_2}(-L_1+\mathcal{C})K_{X_2}) = h^0(X_2,\mathcal{O}(-H))=0$ and $h^2(X_2,\mathcal{O}_{X_2}(L_2-\mathcal{C})) = h^0(X_2,\mathcal{O}_{X_2}(-L_2+\mathcal{C})K_{X_2}) = h^0(X_2,\mathcal{O}(2H))=6$. The two indices are
\begin{eqnarray}\label{IndexForH1App}
  I_{L1} = \int_{X_2} e^{L_1-\mathcal{C}} \mbox{Td}(X_2) = \int_{\mathbb{CP}^2} e^{-2H}\left(1 + \tfrac32 H + H^2 \right) = 0 \:, \\
  I_{L2} = \int_{X_2} e^{L_2-\mathcal{C}} \mbox{Td}(X_2) = \int_{\mathbb{CP}^2} e^{-5H}\left(1 + \tfrac32 H + H^2 \right) = 6 \:.
\end{eqnarray}
This implies that both first homology groups are empty.

\subsubsection*{Example 1 - $\mathcal{C}_{D7\cap E1}$}

This matter curve is given by setting $z_6=z_3=0$ intersected with the equation defining the CY three-fold. Looking at the SR-ideal, we see that we can set $z_2=1$ and $z_5=1$. This fixes two of the scaling relations. We are left with describing the curve by an equation in a two-dimensional toric space:
\begin{align}%
\centering
\begin{tabular}{|c|c|c||c|}
\hline
$z_0$  & $z_1$   & $z_4$  &  $eq_{\mathcal{C}}$ \\
\hline \hline
$1$  & $1$   & $2$   & $6$  \\ 
\hline
\end{tabular}
\label{Curve1126}
\end{align}
First of all, a computation like \eqref{IndexForH1App} shows that the cohomology groups 
$H^1(X_2,\mathcal{O}_{X_2}(L_i-\mathcal{C}))$ are zero also in this case.
Here the computation of the index is trickier to do than before, because the ambient space $X_2$ is singular (the curve generically does not touch the singularity). In any case, we can resolve the two-dimensional toric space and do the computation on the resolved space. 

As before, then, we just need to count holomorphic sections. This can be done already in the singular space. Here $L_1=0$, $L_2=2H$ while $\mathcal{C}=6H$. The first line bundle is trivial and has just one section (the constant), while the sections of the second line bundle are counted by the polynomials of degree $2$, whose number is $4$. Hence,
\begin{eqnarray}
     h^0 (\mathcal{C},[D_D]^{1/2}[D_2]^{1/2} [\cF]) = h^0 (\mathcal{C},\mathcal{O}_{\mathcal{C}}(0)) = 0 - 0 = 0 \:, \\
     h^0 (\mathcal{C},[D_D]^{1/2}[D_2]^{1/2} [\cF]^{-1}) = h^0 (\mathcal{C},\mathcal{O}_{\mathcal{C}}(2H)) = 4 - 0 = 4 \:. 
\end{eqnarray}

\subsubsection*{Example 3 - $\mathcal{C}_{D7\cap E1}$}

This matter curve is given by $z_6=z_0=0$ intersected with the equation defining the CY three-fold. The SR-ideal allows setting $z_4=1$ and $z_5=1$.  We are left with describing the curve by an equation in a two-dimensional toric space:
\begin{align}%
\centering
\begin{tabular}{|c|c|c||c|}
\hline
$z_1$  & $z_2$   & $z_3$  &  $eq_{\mathcal{C}}$ \\
\hline \hline
$1$  & $1$   & $2$   & $2$  \\ 
\hline
\end{tabular}
\end{align}
We see that the equation eliminates $z_3$. We are then left with a $\mathbb{CP}^1$ with coordinates $[z_1,z_2]$. This makes it very easy to count the holomorphic sections. The line bundle restricts to $L_1=-4H$, $L_2=2H$. Hence,
\begin{eqnarray}
     h^0 (\mathcal{C},[D_D]^{1/2}[D_1]^{1/2} [\cF_D]) =  0 \:, \\
     h^0 (\mathcal{C},[D_D]^{1/2}[D_1]^{1/2} [\cF_D]^{-1}) = 3 \:. \\
\end{eqnarray}

\subsubsection*{Example 3 - $\mathcal{C}_{D7\cap E2}$}

Here we have four line bundles of interest, i.e. $L_1=\frac{D_D}{2}+\frac{D_2}{2}+\cF_D-\cF_2$, $L_2=\frac{D_D}{2}+\frac{D_2}{2}-\cF_D+\cF_2$, $L_3=\frac{D_D}{2}+\frac{D_2}{2}+\cF_D+\cF_2$ and $L_4=\frac{D_D}{2}+\frac{D_2}{2}-\cF_D-\cF_2$. 

The intersection curve  is at $z_4=z_6=0$, which allows us to fix $z_0=1$ and $z_2=1$. This curves has the same definition as \eqref{Curve1126}. The four line bundles restrict to $2H$, $0$, $H$ and $H$. This gives respectively $4$, $1$, $2$ and $2$ zero modes.

\subsubsection*{Example 2 - $\mathcal{C}$}

Consider the matter living at the intersection of the brane wrapping $\eta+z_6\psi=0$ with its image $\eta-z_6\psi=0$. The intersection curve is defined by the equations $\eta=\psi=0$ in the CY three-fold $X_3$. The states are counted by the cohomology groups \eqref{ExtD7D7image}, i.e. $H^0(\mathcal{C},\mathcal{O}_{\mathcal{C}}(L)$ with $L=3D_{z_6} \pm 2\cF_D$. Again a lower bound for these numbers is given by $H^0(X_3,\mathcal{O}_{X_3}(L)$, as proven below.  In Example 2, the sections of $\mathcal{O}_{X_3}(L)$ are counted by polynomials of degree $(12,9,5)$ for $L=3D_{z_6} + 2\cF_D$ and by polynomials of degree $(12,9,7)$ for $L=3D_{z_6} - 2\cF_D$. Their number is $301$ and $295$, respectively (remember that for consistency of the Tachyon profile in an orientifold background, we have to count the number of even sections; moreover we have to subtract the number of sections that identically vanish on $X_3$).

\subsubsection*{Bound from sections on the CY}

Remember that $\cC=D_\eta \cap D_\psi$ and $c_1(N_{\cC/X_3})=D_\eta+D_\psi$ ($=7D_{z_6}$). We start from the short exact sequence
\begin{equation}
  0 \rightarrow \mathcal{O}_{X_3}(L-D_\eta) \rightarrow \mathcal{O}_{X_3}(L) \rightarrow \mathcal{O}_{D_\eta}(L) \rightarrow 0 \:.
\end{equation}
Its associated long exact sequence of cohomology groups is
\begin{align}
 0 &  \rightarrow  H^0(X_3,\mathcal{O}_{X_3}(L-D_\eta)) \rightarrow H^0(X_3,\mathcal{O}_{X_3}(L)) \rightarrow H^0(D_\eta,\mathcal{O}_{D_\eta}(L)) \rightarrow  \nonumber \\
  & \rightarrow  H^1(X_3,\mathcal{O}_{X_3}(L-D_\eta)) \rightarrow ...  \:. \nonumber  
\end{align}
In our cases $H^0(X_3,\mathcal{O}_{X_3}(L-D_\eta))=0$, hence $h^0(X_3,\mathcal{O}_{X_3}(L)) \leq h^0(D_\eta,\mathcal{O}_{D_\eta}(L))$ because of injectivity of the map. We now write the short exact sequence for $\cC$ the divisor $D_\psi$ in $D_\eta$:
\begin{equation}
  0 \rightarrow \mathcal{O}_{D_\eta}(L-D_\psi) \rightarrow \mathcal{O}_{D_\eta}(L) \rightarrow \mathcal{O}_{D_\psi}(L) \rightarrow 0 \:,
\end{equation}
Its associated long exact sequence of cohomology groups is
\begin{align}
 0 &  \rightarrow  H^0(D_\eta,\mathcal{O}_{D_\eta}(L-D_\psi)) \rightarrow H^0(D_\eta,\mathcal{O}_{D_\eta}(L)) \rightarrow H^0(\mathcal{C},\mathcal{O}_{\mathcal{C}}(L)) \rightarrow  \nonumber \\
  & \rightarrow  H^1(D_\eta,\mathcal{O}_{D_\eta}(L-D_\psi)) \rightarrow ...  \:. \nonumber  
\end{align}
Luckily $H^0(D_\eta,\mathcal{O}_{D_\eta}(L-D_\psi))=0$, as we prove below. Hence, 
\begin{equation}
h^0(\mathcal{C},\mathcal{O}_{\mathcal{C}}(L))\geq h^0(D_\eta,\mathcal{O}_{D_\eta}(L))  \geq h^0(X_3,\mathcal{O}_{X_3}(L)) \:.
\end{equation}

We finish by proving $H^0(D_\eta,\mathcal{O}_{D_\eta}(L-D_\psi))=0$.
By the same steps as above, we write the long exact sequence
\begin{align}
 0 &  \rightarrow  H^0(X_3,\mathcal{O}_{X_3}(L-D_\psi-D_\eta)) \rightarrow H^0(X_3,\mathcal{O}_{D_\eta}(L-D1)) \rightarrow H^0(D_\eta,\mathcal{O}_{\mathcal{C}}(L-D_\psi)) \rightarrow  \nonumber \\
  & \rightarrow  H^1(X_3,\mathcal{O}_{X_3}(L-D_\psi-D_\eta)) \rightarrow ...  \:. \nonumber  
\end{align}
We have $H^0(X_3,\mathcal{O}_{D_\eta}(L-D1))=0$ (by counting holomorphic sections on $X_3$). Moreover (using Serre duality) $H^1(X_3,\mathcal{O}_{X_3}(L-D_\psi-D_\eta)) =H^2(X_3,\mathcal{O}_{X_3}(-L+D_\psi+D_\eta)) =0$, because the divisor $-L+D_\psi+D_\eta$ is ample in $X_3$ (this can be checked, by seeing that it lies inside the K\"ahler cone, for Example 1). This implies $H^0(D_\eta,\mathcal{O}_{D_\eta}(L-D_\psi))=0$ as we wanted.

\bibliographystyle{utphys}
\bibliography{myrefs}

\end{document}